\documentclass[fleqn,10pt,dvipsnames]{wlscirep}
\usepackage[utf8]{inputenc}
\usepackage[T1]{fontenc}
\usepackage{bm}
\usepackage{comment}
\usepackage{amsmath,amssymb,physics}
\usepackage{ulem}
\usepackage{dsfont}
\usepackage{wrapfig}
\usepackage{mathtools}
\usepackage{subcaption}
\usepackage{tikz}
\usepackage[framemethod=TikZ]{mdframed}
\usepackage{braket}
\usepackage[mathscr]{eucal}
\usepackage[most]{tcolorbox}

\DeclareMathAlphabet{\mathcal}{OMS}{cmsy}{m}{n}

\newtcolorbox{calloutbox}[1][]{
colback=gray!10,    
colframe=gray!50,   
sharp corners,      
boxrule=0.5pt,      
arc=0mm,            
left=1em,           
right=1em,          
top=1em,            
bottom=1em,         
fontupper=\itshape, 
#1                  
}

\usetikzlibrary{3d, decorations.pathmorphing}
\tikzset{snakeX/.style={decorate, decoration={snake, amplitude=.6mm, segment length=3mm}}}
\tikzset{snakeY/.style={decorate, decoration={snake, amplitude=.6mm, segment length=5mm}}}
\tikzset{snakeZ/.style={decorate, decoration={snake, amplitude=.6mm, segment length=7mm}}}

\tikzset{X/.style={color=myred, line width=1.5mm, snakeX}}
\tikzset{Y/.style={color=LimeGreen, line width=1.7mm}}
\tikzset{Z/.style={color=MidnightBlue, line width=1.7mm, snakeZ}}

\definecolor{myred}{RGB}{255, 79, 79}

\newcommand{\fh}{H} 
\newcommand{\qh}{\widetilde{H}} 

\newcommand{\al}[1]{\begin{align}#1\end{align}}

\newcommand{\g}{\gamma}
\newcommand{\gb}{\overline{\gamma}}

\newcommand{\sel}{\text{\footnotesize \textsf{SEL}}}
\newcommand{\prep}{\text{\footnotesize \textsf{PREP}}}
\newcommand{\swap}{\text{\footnotesize \textsf{SWAP}}}
\newcommand{\fswap}{\text{\footnotesize \textsf{fSWAP}}}
\newcommand{\cz}{\text{\footnotesize \textsf{CZ}}}
\newcommand{\cnot}{\text{\footnotesize \textsf{CNOT}}}

\newcommand{\ketf}[1]{|{#1})}

\newcommand{\arxiv}[1]{
}

\definecolor{CommentGreen}{RGB}{5,130,5}
\definecolor{CommentRed}{RGB}{220,50,25}

\title{Putting fermions onto a digital quantum computer} 

\author[1,*]{Riley W. Chien}
\affil[1]{Sandia National Laboratories, Albuquerque, NM 87185, USA}

\author[2]{Mitchell L. Chiew}
\affil[2]{DAMTP, Centre for Mathematical Sciences, University of Cambridge, Cambridge CB3 0WA, UK}

\author[3]{Brent Harrison}
\author[3]{Jason Necaise}
\author[3]{Weishi Wang}
\affil[3]{Department of Physics and Astronomy, Dartmouth College, Hanover, NH 03755, USA}

\author[4]{Maryam Mudassar}
\affil[4]{Joint Center for Quantum Information and Computer Science, University of Maryland, College Park, MD 20742, USA}

\author[5]{Campbell McLauchlan}
\affil[5]{School of Physics, University of Sydney, Sydney, NSW 2006, Australia}

\author[6,7]{Thomas M. Henderson}
\author[6,7]{Gustavo E. Scuseria}
\affil[6]{Department of Chemistry, Rice University, Houston, TX 77005-1892, USA}
\affil[7]{Department of Physics and Astronomy, Rice University, Houston, TX 77005, USA}

\author[8]{Sergii Strelchuk}
\affil[8]{Department of Computer Science, University of Oxford, Oxford OX1 3QG, UK}

\author[3,$\dagger$]{James D. Whitfield}
\affil[*]{e-mail: rwchien@sandia.gov}

\affil[$\dagger$]{e-mail: james.d.whitfield@dartmouth.edu}

\begin{abstract}
Quantum computers are expected to become a powerful tool for studying physical quantum systems. Consequently, a number of quantum algorithms for studying the physical properties of such systems have been developed. While qubit-based quantum computers are naturally suited to the study of spin-1/2 systems, systems containing other degrees of freedom must first be encoded into qubits. Transformations to and from fermionic degrees of freedom have long been an important tool in physics and, now the simulation of fermionic systems on quantum computers based on qubits provides yet another application. In this perspective, we review methods for encoding fermionic degrees of freedom into qubits and attempt to dispel the persistent notion that fermionic systems beyond one dimension are fundamentally more difficult to deal with. 
\end{abstract}

\begin{document}
\maketitle
\thispagestyle{empty}

\begin{calloutbox}
\noindent \textbf{Key points:} 

\begin{itemize}
    \item Physical systems have complex interactions that can involve fermions and computing physically relevant quantities is classically challenging; quantum simulation algorithms for digitally preparing and estimating observables are actively researched.
    \item The study of physical systems beyond the reach of classical methods has been identified as one of the most important applications of the emerging technology.
    \item Simulating fermions on a quantum computer requires encoding the antisymmetric exchange into qubits, the basic memory elements of most quantum computers.
    \item First quantization, where the number of fermions and their anti-symmetrization correlation are explicitly encoded in the many-particle states, provides a compact representation of many-electron systems restricted to a fixed particle-number subspace.
    \item Second quantization, a generically applicable formalism natural in many applications areas, is amenable to a variety of encoding methods depending on the structure of the problem and available computing resources.
\end{itemize}
\end{calloutbox}

\pagebreak

\flushbottom
\tableofcontents

\section{Introduction}

The basic goal of computational quantum mechanics is to accurately determine the properties of systems of many interacting particles.  We may, for example, wish to compute the vibrational spectrum of a molecule, a correlation function in a lattice model, or the quadrupole moment of a nucleus.  When interactions between particles are weak, this is relatively straightforward, and mean-field methods which approximate the interaction between particles can provide reasonable estimates. In general, however, more sophisticated computational techniques are required. Ultimately, for some problems, the tools provided by classical computation may not be able to efficiently produce accurate enough results.

About forty years ago, Manin\cite{manin1980computable} and Feynman\cite{feynman2018simulating} suggested that quantum computers are the natural platform for simulating quantum systems. Although large-scale fully error-corrected quantum computers are not yet online, substantial experimental progress has been made towards making devices with more physical qubits and with lower error rates. Demonstrations of noisy quantum simulations of physical systems have been conducted for years \cite{peruzzo2014variational,arute2020observation,farrell2024scalable}, but recently error correction primitives have been demonstrated in a number of platforms\cite{google2023suppressing,bluvstein2024logical,paetznick2024demonstration}. Theoretical work in preparation for the deployment of large scale quantum computers has been ongoing in parallel to the experimental progress. This includes algorithm development, resource estimation, and comprehensive protocols for computing observables of commercial and scientific interest.

Physical systems involving fermionic particles are common targets for simulation on quantum computers, due to their centrality across quantum chemistry, material science, and high-energy physics\cite{abrams_simulation_1997,ortiz2001quantum,lanyon2010towards,wecker2014gate,reiher2017elucidating}. However, there is an important challenge when studying fermions on standard quantum computing architectures. The fundamental units in most digital quantum computers are qubits, objects carrying two-level Hilbert spaces which compose with a tensor product. Fermionic systems on the other hand posses a subtle nonlocality stemming from antisymmetric exchange that is not present for spins. This means that before we can treat fermionic systems on a digital quantum computer, we must write them in the language of qubits. Such \textit{fermion-to-qubit mappings} or \textit{fermionic encodings} will be the topic of this review.

To place these fermion-to-qubit mappings into context, we will first survey several major application domains that motivate fermionic simulations on quantum hardware. In particular, we discuss how quantum chemistry, condensed matter physics, and high-energy physics serve as important targets for quantum simulation, each involving fermionic degrees of freedom that pose significant challenges for classical computation. We will then outline the essential quantum algorithms and techniques underpinning these simulations, including methods for state preparation of relevant fermionic states, approaches for observable estimation to extract physical quantities, and illustrative algorithmic primitives that serve as building blocks for more complex quantum simulation protocols. This background provides the necessary context and motivation for the technical developments that follow. We then discuss the challenge of representing fermionic systems on a quantum computer. We introduce and compare the two principal frameworks for encoding fermionic models – first quantization and second quantization. They are fundamentally different in the way they treat fermionic statistics. In the second-quantized setting, we further review several important fermion-to-qubit encoding strategies. In particular, we examine the conventional Jordan–Wigner transformation alongside more recently proposed schemes, including alternative ancilla-free encodings, symmetry-based qubit reduction techniques, and local encodings. For each encoding approach, we discuss the associated resource requirements and trade-offs in implementation, providing guidance on how to choose an appropriate mapping for a given simulation task. Together, these sections offer a comprehensive roadmap of the techniques, application areas and design considerations when simulating fermionic systems on a digital quantum computer.

\section{Applications}
Decades of development of classical algorithms for simulating quantum systems have yielded a number of widely-used techniques. Often these algorithms rely on some assumption about the physical system in question, and their success depends on the validity of these assumptions.  The hope is that quantum algorithms will enable us to solve a broader class of problems than we have been able to study thus far.  Here, we highlight some of the most important targets of quantum simulation.

\subsection{Quantum chemistry}
Quantum chemistry uses the Schr\"odinger equation to predict the properties and reactions of molecules.  While the quantum effects of the electrons and nuclei in molecular systems can be treated together, one often makes the Born--Oppenheimer approximation, yielding a purely electronic Hamiltonian that depends parametrically on the nuclear coordinates $\{\bm{R}_A\}_{A}$: 
\begin{equation}\label{eq:Helec}
    H_{\mathrm{elec}} = \sum_{i=1}^{\eta}\Big(\underbrace{-\frac{1}{2}\nabla_i^2}_{T_i} \,+\, V_i(\bm{r}_i;\,\{\bm{R}_A\}_{A}) + \sum_{j=i+1}^{\eta}\underbrace{\frac{1}{\lVert \bm{r}_i - \bm{r}_j \rVert_2}}_{U_{ij}}\Big)\, .
\end{equation}

In practice, the Hamiltonian is discretized via the Galerkin method with respect to a chosen single-particle basis set $\{\varphi_{m}(\bm{r},\omega)\}_{m=1}^{M}$ where $\varphi_{m}(\bm{r},\omega) = \phi_m(\bm{r})\sigma_m(\omega)$ is the wavefunction of a single electron, i.e., a spin-orbital. Specifically, $\phi_m(\bm{r})$ is referred to as a spatial orbital, and $\sigma_m(\omega) \in \{\alpha(\omega),\,\beta(\omega)\}$ is one of the two spin-{1/2} eigenfunctions. Correspondingly, the matrix elements of a discretized one-body operator $O$ such as $T_i$ and $V_i$ are computed through the one-body integrals 
\begin{align}
    O_{ij} 
    &= \int d\bm{r}\,\, 
    \phi^*_{i}(\bm{r}) 
    \,O\, 
    \phi_{j}
    (\bm{r})  \delta_{\sigma_i \sigma_j}.
    \label{eq:onebody_integral}
\end{align}
Similarly, the integrals for the tensor elements of the discretized electron-electron Coulomb operator are defined as
\begin{align}
    U_{ijkl} 
    & = \iint d\bm{r}_1 d\bm{r}_2 \,\, \phi^*_{i}(\bm{r}_1)\phi^*_{j}(\bm{r}_2) \frac{1}{\lVert \bm{r}_1 - \bm{r}_2 \rVert_2} \phi_{k}(\bm{r}_1)\phi_{l}(\bm{r}_2)\delta_{\sigma_i\sigma_k} \delta_{\sigma_j\sigma_l}\, ,
    \label{eq:twobody_integral}
\end{align}
with suitable modifications should the operators act on spin coordinates as well as spatial coordinates.  These same integrals are used when working in second quantization, in which the Hamiltonian is re-expressed as
\begin{equation}
H_\mathrm{elec} = \sum_{i,j}^{M} (T_{ij} + V_{ij})  a_i^\dagger  a_j + \frac{1}{2}  \sum_{i,j,k,l}^{M} U_{ijkl}  a_i^\dagger  a_j^\dagger  a_l  a_k
\end{equation}
where orbital $\varphi_m(\bm{r},\omega)$ is associated with the second-quantized mode created by $a_m^\dagger$. Once we represent the electronic Hamiltonian discretized by a basis set in the second quantization formalism, $H_\mathrm{elec}$ is no longer characterized by $\eta$. Instead, it contains eigenstates of arbitrary particle numbers up to $M$. Given a discretized electronic Hamiltonian, electronic structure theory solves for stationary states, while electronic dynamics focuses on obtaining time-dependent states.  The stationary states provided by electronic structure theory yield potential energy surfaces on which \textit{ab initio} molecular dynamics can be studied.

Many algorithms have been developed for electronic structure\cite{AspuruGuzik2005,whitfield2011simulation,cao2019quantum,bauer2020quantum,mcardle2020quantum,motta2022emerging} and dynamics\cite{Kassal2008,Sawaya2019,Ollitrault2020,Miessen2023}. These algorithms frequently use a mixture of quantum and classical computation, but in the longer term purely quantum simulations are possible. There have also been numerous demonstrations of simulations of small molecules on quantum devices\cite{OMalley2016,Kandala2017,Colless2018,Hempel2018,Google2020,navickas2024experimentalquantumsimulationchemical}.

\subsection{Condensed matter physics}

Much of condensed matter physics and material science is concerned with understanding electrons in a periodic potential. We might be interested in real materials, in which case the Hamiltonian is the same as Eq. \ref{eq:Helec}, or in model systems that provide insight into the qualitative behavior of a class of such materials. Whether studying real materials or model Hamiltonians, one might be interested in computing properties such as lattice constants, band structures, and so on, or in understanding the phase diagram and order parameters.

Model systems have been a target of early quantum simulations and have guided algorithm development due to being simple to formulate but challenging to study numerically. The Fermi--Hubbard model, a model of spinful fermions hopping on a lattice with a term penalizing doubly-occupied sites\cite{Hubbard1963},
\begin{align}
\fh = -t\sum_{\mathbf{x},\hat{e}}\sum_{\sigma=\{\uparrow,\downarrow\}} (a^{\dag}_{\mathbf{x}+\hat{e},\sigma}a_{\mathbf{x},\sigma} + a^{\dag}_{\mathbf{x},\sigma}a_{\mathbf{x}+\hat{e},\sigma}) + U\sum_{\mathbf{x}} a^{\dag}_{\mathbf{x},\uparrow} a_{\mathbf{x},\uparrow}a^{\dag}_{\mathbf{x},\downarrow} a_{\mathbf{x},\downarrow}\, ,
\label{eq:hubbard}
\end{align}
is a paradigmatic example of a strongly-interacting condensed matter system. For strong Hubbard repulsion $U$, and physically-relevant filling fractions and lattice geometries, this model displays a wealth of different low temperature phases, making it very challenging to study.

There has been a lot of progress designing algorithms for studying quantum simulations of condensed matter systems\cite{babbush2018low}. For example, protocols to simulate particular model systems such as the Hubbard model have been devised\cite{wecker2015solving,cade2020fermihubbard}. The study of realistic material Hamiltonians has also been extensive\cite{bauer2016hybrid,rubin2023fault,clinton2024towards,berry2024quantum}. Algorithms have been formulated in both first and second quantization as well as in real space and momentum space.

\subsection{High-energy physics}

The Standard Model (SM) of particle physics, a quantum field theory (QFT) of non-abelian gauge fields coupled to fermionic and scalar fields, stands as our most comprehensive framework for understanding the fundamental forces that govern the universe. Many mysteries within the Standard Model remain to be understood, such as confinement and the computation of particle masses. In addition, the standard model is known to be incomplete, as it fails to account for key phenomena such as dark matter, the matter-antimatter imbalance, and the masses of neutrinos. These unresolved issues highlight the necessity for theories that extend beyond the Standard Model. 

For strongly-coupled theories, such as the quantum chromodynamics (QCD) sector of the SM, numerical study is one of the few ways to make progress. This requires discretizing the theory into a spacetime lattice field theory, in particular a lattice gauge theory. Despite significant progress in lattice gauge theories and their simulation on high-performance computers, simulating QFTs on classical computers poses substantial challenges due to the computational complexity involved, especially in non-perturbative regimes. This remains true for the efforts of QCD, making both the computation and the interpretation of results challenging.

The implementation of quantum algorithms for strongly-interacting lattice QFTs remains one of the promising areas for quantum computation. In the context of quantum simulation, it is most common to work in the Hamiltonian formulation, where space is discrete, but time remains continuous. There have been promising early results on the development of quantum algorithms that can efficiently compute relativistic scattering probabilities in scalar $\phi^4$ theory in various spacetime dimensions\cite{jordan2011quantum,jordan2012quantum,jordan2018bqp}, as well as fermionic theories\cite{jordan2014quantum}. In the past decade, there has been a lot of work on quantum simulation of Hamiltonian gauge theories\cite{lamm2019general,kan2021lattice,tong2022provably,bauer2023quantum,bauer2023quantum_natrevphys,lamm2024block,rhodes2024exponential} as well as effective field theories\cite{watson2023quantum}.

The Kogut-Susskind formulation gives a by-now standard Hamiltonian formulation of lattice gauge theories \cite{kogut1975hamiltonian}. This requires representing both the gauge field degrees of freedom as well as the fermionic matter transforming in a representation of the gauge group, both of which must be addressed. Note that when putting field theories onto a lattice there are also issues stemming from chirality\cite{nielsen1981absence} which we do not address as we consider these to come before encoding onto a quantum computer. 

As a stepping-stone to more complicated theories, one notable model that has been explored is the Schwinger model\cite{schwinger1962gauge}. This quantum field theory, defined in 1+1 dimensions, describes the interaction between fermions and a $U(1)$ gauge field, providing a useful testing ground for quantum simulations of Hamiltonian lattice gauge theories\cite{kuhn2014quantum,farrell2024scalable}. Insights gained from simulating the Schwinger model have helped develop computational methods that can be applied to more complex scenarios, setting the stage for extending these techniques to other gauge theories, including QCD\cite{alexandru2019gluon,davoudi2021search,kan2021lattice,lamm2024block,rhodes2024exponential}.

\begin{figure}
    \centering
    \includegraphics[width=0.7\linewidth]{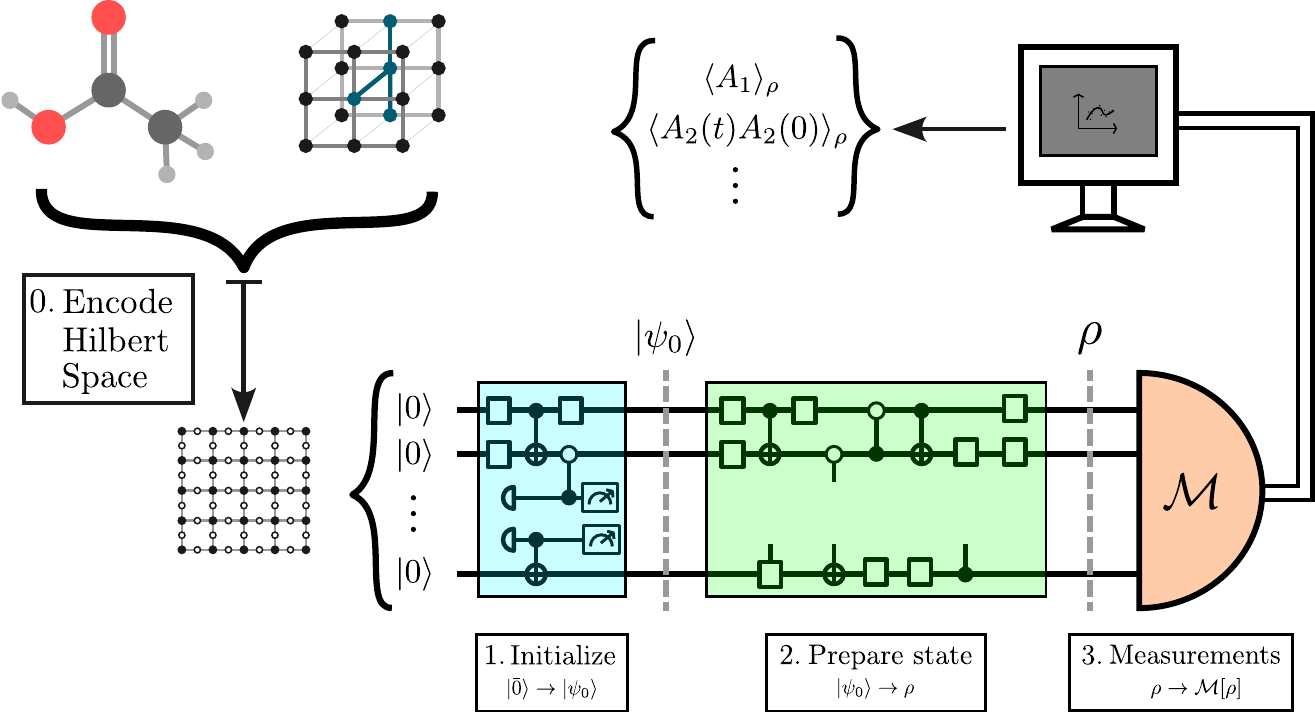}
    \caption{Simulation overview through four steps: 0. Choose a representation 1. Initialize the easy-to-prepare physical starting state $|\psi_0\rangle$ 2. Prepare the desired state of interest $\rho$. 3. Perform measurements $\rho\to\mathcal{M}[\rho]$ and estimate observables. }
    \label{fig:simulation_overview}
\end{figure}

\section{Relevant algorithms}
The interest in using quantum computers for computational physics and chemistry is due to the hope that quantum computers will be able to offer a dramatic speed up over classical computers for physically relevant problems. They will hopefully allow us to perform \textit{ab initio} calculations of physical systems out of the reach of conventional computing systems. Ultimately, this depends on the existence of algorithms for the problems of interest: in this case, problems involving systems of fermions. In order to introduce some of the methods used in the quantum algorithms literature, we will walk through a hypothetical protocol to compute the average of some observables $\{\mathcal{O}_i\}$ in some desired state $\rho$. This is purposely generic and we will describe the different considerations when the observables are either static, equilibrium properties of the system or dynamic, multi-time correlators. We will also consider the different methods of preparing the desired state $\rho$. Along the way, we will mention the theoretical hardness of the various tasks and the quantum resources required.

\subsection{Choice of representation and initialization}

The majority of quantum algorithms expect a formulation of a problem in a standard qubit representation and are agnostic to whether the physical system being studied is fermionic, bosonic, or otherwise. To utilize general-purpose quantum algorithms for fermionic systems, we must translate our problem into qubit form. This requires a dictionary that maps between fermionic and multi-qubit states and operators. The choice of representation can have a profound impact on the character of the simulation, with different representations sometimes leading to very different resource requirements. We will spend much of the rest of this review discussing such representations, the choices involved, and how it impacts the resources required for quantum simulations. For some encodings, such as first quantization and local second-quantized encodings, the computational basis states may not correspond to physical states of the fermionic system under study. In these cases, one must utilize an additional subroutine to initialize into a state $|\psi_0\rangle$ in the physical subspace.

\subsection{State preparation}
Preparing physically relevant states -- ground states, low-energy excited states, or thermal states -- is computationally hard for generic Hamiltonians. This can be simplified by taking advantage of some additional structure in the problem. There are many proposed approaches, but we will highlight three: adiabatic state preparation, phase estimation, and engineered dissipation. 

In adiabatic state preparation, one places a parent Hamiltonian $\qh_0$ as the initial point and the Hamiltonian $\qh$ whose ground state we wish to prepare as the end point of a one-parameter family of Hamiltonians, $\qh(t) = (1-s(t))\qh_0 + s(t) \qh$ \cite{farhi2000quantum,albash2018adiabatic}. As long as the spectral gap between the ground state and first excited state remains open for all values of $t$, we can prepare the ground state of $\qh$ by simulating the time-dependent Hamiltonian evolution with an error that depends on the gap and the rate $ds/dt$ \cite{kato1950adiabatic}. The key assumption is that we can find a path through the space of Hamiltonians throughout which the gap remains open. 

In chemistry, there are established classical methods which can be incorporated in quantum state preparation algorithms to produce trial states that have nonvanishing overlaps with the ground state\cite{lee2023evaluating}. Measuring such trial states in the energy eigenbasis via phase estimation will probabilistically project onto the desired ground state\cite{kitaev1995quantum}. 

In the lab, physical systems under study are never truly isolated from the environment, but rather evolve under open system dynamics,
such as the evolution $\partial_t \rho = \mathcal{L}[\rho]$ generated by
a Lindbladian $\mathcal{L}$. In contact with a thermal bath at inverse temperature $\beta$, typical quantum states reach a fixed point of the open system dynamics which is the thermal state $\lim_{t\to\infty} \exp{t\mathcal{L}}[\rho_0] \propto \exp \{-\beta \qh\}$. By analogy, one can engineer discrete- or continuous-time open system dynamics with the desired state as a fixed point of the evolution~\cite{temme2011quantum,chen2023quantum,chen2023efficient,cubitt2023dissipative,zhang2023dissipative}. The efficiency of such approaches then relies on the efficiency of simulating the engineered dissipative dynamics and time-scale with which the dynamics converges to its fixed point.

These different approaches use some of the same algorithmic primitives that are standard in quantum algorithms, such as Hamiltonian simulation, phase estimation, and the quantum Fourier transform. Which of these approaches is preferred in a simulation may depend on the observables one wishes to compute and prior knowledge about the system under study.

\subsection{Observable estimation}

Once the desired state $\rho$ is prepared, we can proceed to measure the observable whose average value we wish to compute. The phase estimation algorithm\cite{kitaev1995quantum} is a general method for measuring observables with initially unknown spectra. For unitary observables such as Pauli operators with the known spectrum $\{+1,-1\}$, a simplified 1-bit version of phase estimation known as the Hadamard test requires a single application of the unitary operator conditioned on an ancilla. These methods straightforwardly extend to estimating multi-time observables such as Green's functions $\langle A(t)B\rangle_{\rho}$. We can use a further simplified strategy for estimating averages of observables which are given as sums of Pauli terms, $\langle \qh \rangle_{\rho}=\sum_i g_i \langle P_i\rangle_{\rho}$. To get a sample for a single Pauli term $P_i$, we can simply measure each qubit in its support in the basis of the operator's action and multiply the $\pm 1$ outcomes for each factor such that the average gives an estimate of $\langle P_i\rangle$.

An altogether different measurement paradigm is that of the classical shadows family of protocols\cite{huang2020predicting,elben2023randomized}. In these techniques, each qubit is measured after $\rho$ has been acted on by a unitary randomly drawn from an ensemble. Under various conditions, the measurement channel is invertible and if the unitaries are efficiently simulatable classically, the measurement outcomes along with the selected unitaries can be used to construct an estimator for various observables. In this framework, a large number of observables can be accurately estimated after measuring only a small number of copies of the prepared state $\rho$. Which types of observables are estimable from the shadows depends on the ensemble of unitaries utilized. Random single-qubit Cliffords are a natural choice that allows one to efficiently estimate many few-body observables, while global Cliffords allow one to estimate many-body observables. Variants of the classical shadows framework of randomized measurement have been proposed that utilize ensembles of unitaries that are inherently fermionic in nature. These include the group of Majorana $\swap$ operators\cite{zhao2021fermionic}, free-fermion or matchgate unitaries \cite{wan2023matchgate}, particle number-preserving unitaries \cite{low2022classical}, and multi-copy protocols\cite{king2024triply}.

\subsection{Illustrative algorithmic primitives}\label{sec:algorithm} 

\begin{figure}
    \centering
    \includegraphics[width=0.8\linewidth]{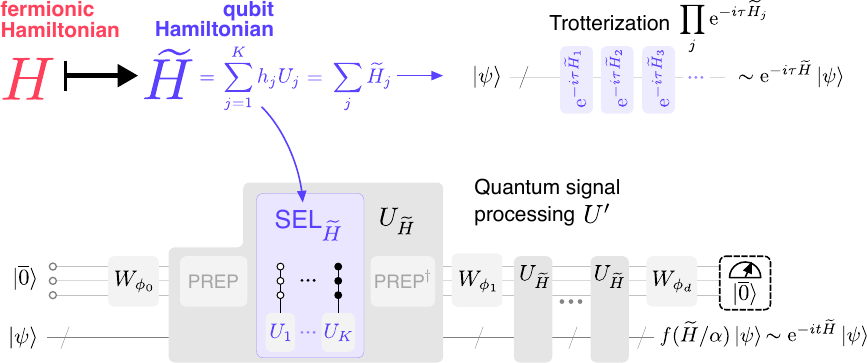}
    \caption{Some Hamiltonian simulation algorithms applied to fermionic systems. The fermion-to-qubit mapping determines the qubit Hamiltonian, which in turn determines the resource cost of simulation algorithms.
    }
    \label{fig:algorithms}
\end{figure}

Because Hamiltonian simulation is so ubiquitous in quantum algorithms, we will focus more in-depth on the two most common methods for simulating time evolution: Trotterization and quantum signal processing.  When choosing a fermionic representation for the quantum computer (the topic of Sec.\ \ref{sec:representation}), one should have in mind these methods of Hamiltonian simulation and their efficiency.

\textit{Trotterization:}
Product formulae such as the first-order Trotter-Suzuki formula approximate the exponential of $\qh = \sum_{j} \qh_j$, which may be difficult to implement \textit{a priori}, via a product of exponentials of each term which may be straightforward to implement:
\begin{equation}
    \text{\textbf{Trotterization}:} \quad \prod_{j} \mathrm{e}^{-\mathrm{i}\tau\qh_j}  = \mathrm{e}^{-\mathrm{i}\tau\qh} + \mathcal{O}(\tau^2)\, .
    \label{eq:trotter}
\end{equation}
In a standard Pauli decomposition of the Hamiltonian, the Trotterized time evolution is approximated by a sequence of Pauli rotations that can be compiled into Clifford gates and single-qubit rotations. For total simulation time $T$ and an error tolerance $\epsilon$, one needs to divide $T$ into $r=\mathcal{O}(T^2/\epsilon)$ Trotter steps each of length $T/r$\cite{lloyd1996universal}. Higher-order Trotter formulae can improve this scaling to $r=\mathcal{O}(T^{1+\frac{1}{n}}/\epsilon^{\frac{1}{n}})$ at the cost of increasingly complex steps with order $n$\cite{suzuki1991general}. Despite the suboptimal scaling in the worst-case, low-order product formulae remain
relevant due to their simplicity, requiring no ancillary qubits or complicated subroutines\cite{bosse2025efficient}. Additionally, system-dependent error analysis can often show that the naive error scaling is overly pessimistic\cite{childs2019nearly,childs2021theory} and further, product formula methods often perform much better empirically than provable error bounds suggest.

\textit{Quantum signal processing:}
A flexible algorithmic framework which has many applications such as Hamiltonian simulation is that of quantum signal processing (QSP) and its generalizations\cite{low2017optimal,low2019hamiltonian,gilyen2019quantum,motlagh2024generalized}. In this framework, one utilizes a \textit{block encoding} of a matrix, such as the encoded Hamiltonian $\qh$, in which the (not necessarily unitary) matrix is embedded into the top left block of a unitary matrix. In order for the embedding to be possible, the operator norm of the embedded matrix must be at most 1. To ensure this, we can obtain an upper bound on the norm, $\|\qh\|\leq \alpha$, and block encode the rescaled matrix $\qh/\alpha$ into the $(\bar{0},\bar{0})$-block of the ancilla register:
\al{
 \text{\textbf{Block encoding:}} \quad   (\langle \bar{0}| \otimes I ) U_{\qh} (|\bar{0}\rangle \otimes I) = \qh/\alpha\, .
}
Interleaving $U_{\qh}$ with rotations $\{W_j\}_{j=0}^{d}$ acting on the ancilla register allows one to implement polynomial functions of the block encoded matrix:
\al{
    \text{\textbf{Quantum signal processing:}} \quad (\langle \bar{0}| \otimes I ) W_0  \left( \prod_{j=1}^d  U_{\qh} W_j \right) (|\bar{0}\rangle \otimes I) = f(\qh/\alpha)\, .
}
Using quantum signal processing for polynomial approximations of desired functions provides
algorithms for implementing matrix functions on states, a format which encapsulates the solution of many tasks.

The complexity of the algorithm then depends on the cost of implementing the block encoding matrix, $U_{\qh}$, and the number times it must be implemented, which is given by the degree of the desired polynomial. The complexity also depends on the amplitude of the $|\bar{0}\rangle$ state in the output of the ancilla register as this determines the success probability of the block encoding. One can post-select on the ancilla state being measured in the $|\bar{0}\rangle$ state or use amplitude amplification to boost the probability towards one. For Hamiltonian simulation, we want to approximate the matrix exponential, $f(\qh/\alpha) \approx e^{i t \qh}$. For error tolerance $\epsilon$, this can be achieved with $\mathcal{O}(t + \log(1/\epsilon))$ applications of $U_{\qh}$, the optimal scaling\cite{low2017optimal,low2019hamiltonian}. 

We still need to actually construct the block encoding unitary $U_{\qh}$. A common framework for constructing block encodings is the linear combination of unitaries (LCU)\cite{childs2012hamiltonian}. One expresses the Hamiltonian as a linear combination of unitary operators, $\qh = \sum_{j=1}^K h_j U_j$. The sum of the coefficients gives an upper bound to the norm, $\|\qh\|\leq \alpha =\sum_j^K |h_j|$ and serves as the normalization factor. One breaks $U_{\qh}$ up into a composition of two subroutines called prepare and select. The prepare circuit, $\prep$ acting on the ancilla register of $\lceil \log_2 K\rceil$ qubits is constructed to encode the coefficients of the Hamiltonian terms, before the select circuit, $\sel_{\widetilde{H}}$, conditionally applies the unitaries in the decomposition of $\qh$. The LCU block encoding is then constructed from the composition
\begin{equation}
    \text{\textbf{LCU block encoding:}} \quad \prep |\bar{0}\rangle = \sum_{j} \sqrt{\frac{h_j}{\alpha}} |j\rangle\, , \quad  \sel_{\qh} = \sum_j |j\rangle \! \langle j| \otimes U_j \quad \implies \quad U_{\qh} = (\prep^\dag \otimes I)\cdot \sel_{\qh}\cdot (\prep\otimes I)\, .
    \label{eq:prepare}
\end{equation}
The cost of $\prep$ depends on the coefficients $h_j$ being encoded; 
structure such as repeated coefficients can reduce this cost. The LCU form of the Hamiltonian is particularly simple for Hamiltonians originating from second-quantized fermionic systems, where a Pauli decomposition is readily obtained. Methods for constructing such oracles include binary indexing, which requires a logarithmic number of ancillas in the number of terms, and unary iteration which requires a linear number of ancillas but has superior non-Clifford gate scaling \cite{babbush2018encoding}. Alternatively to the linear combination of unitaries, efficient block encodings can also be constructed for sparse matrices\cite{berry2009black}. 

Hamiltonian simulation by QSP is capable of simulating Hamiltonian evolution with optimal complexity, however there are overheads involved which are not a factor in product formula time evolution such as the compilation of the $\prep$ circuit. So while the scaling for QSP may be better than Trotterization, for a particular system, the total costs may not and one will have to check which method is preferable. 

\section{Representing fermionic systems on a digital quantum computer}\label{sec:representation} 
In order to simulate any physical system, we need to formulate the problem in a language that our computer can understand.  In other words, the computer must be able to hold a description of the system in its memory upon which it can perform operations. On a digital quantum computer, the relevant physical states and operators must be represented within the multi-qubit Hilbert space of the quantum computer. For some degrees of freedom, approximations may be required, such as when the Hilbert space of the system of interest is infinite-dimensional, in which case an approximation error will be incurred. Ultimately, we would like to have a dictionary between any physical, calculable quantity in the target system and quantities we can calculate on the quantum computer.

In the case of fermionic systems, there are two familiar ways to represent the system on a digital quantum computer consisting of qubits: first and second quantization. The two representations differ in whether the fermionic nature of the system manifests in the states or the operators. In first quantization, the system is described by a many-body state which is explicitly antisymmetric upon permuting the particle registers. In second quantization, the antisymmetry is manifest in the anticommutation relations of the fermionic operators. The choice of qubit representation is determined in part by the formulation of the problem one wants to solve. 
In this section, we provide an overview of both the first- and second-quantized approaches. In Sec.\ \ref{sec:comparison}, 
we lay out potentially differentiating cost factors between the quantization approaches when employed in fermionic simulation algorithms.

Before discussing these two different representations of fermions, it is important to mention that many physical systems we would like to study contain other types of degrees of freedom in addition to fermions. In particular, there may be bosonic modes, such as vibrational modes or latticized bosonic fields; there may also be lattice gauge fields in which case the states of the system will be subjected to constraints imposed by gauge invariance. In some cases, the choice of truncation of infinite-dimensional degrees of freedom will be crucial to the accuracy of the simulation.

\subsection{First quantization}\label{sec:firstQ}

When dealing with a given system with a fixed number of particles $\eta$, it can be natural to construct operators in terms of actions on the states of each particle. In the fermionic case, this requires a restriction to an antisymmetric state subspace. This method of construction is widely used in the quantum chemistry literature, as it is particularly common in the particle-preserving electronic structure problem. 

\begin{figure}[t]
    \centering
    \includegraphics[width=\linewidth]{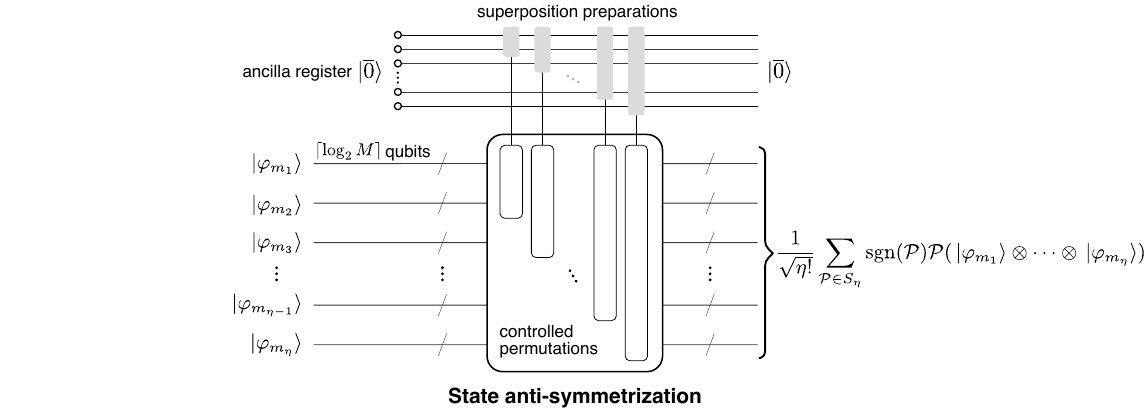}
    \caption{The typical form for state anti-symmetrization as part of fermionic simulation in first quantization. To represent a fermionic state, the circuit input is a product of qubit states $|\varphi_{m_i}\rangle$ which indicate that the $i$th fermion is in the spin-orbital that has label $m_i$. The output of the circuit is an antisymmetrized qubit representation of the desired fermionic state.} 
    \label{fig:1stquant}
\end{figure}

On a quantum computer, the first-quantized representation allocates a register of $\lceil \log_2 M \rceil$ qubits for each particle. Each qubit register has a dimension equal to the number of spin-orbitals ($M$) used to discretize the electronic Hamiltonian, cf.  Eq.\ (\ref{eq:onebody_integral}, \ref{eq:twobody_integral}). Correspondingly, the computational basis states of each qubit register $|\varphi_{m_i}\rangle$ specify (e.g., in binary) which spin-orbital the particle occupies, with $m_i\in\{1,\ldots,M\}$ and $i \in \{1, \dots , \eta\}$. The fermionic nature of the particles mandates that the multi-particle basis states, the Slater determinant states, must be anti-symmetrized,
\begin{equation}
    \text{\textbf{State anti-symmetrization:}} \quad \ket{\varphi_{m_1}} \otimes \dots \otimes \ket{\varphi_{m_\eta}} \longmapsto
    \frac{1}{\sqrt{\eta!}}\sum_{\mathcal{P}\in S_{\eta}}\mathrm{sgn}(\mathcal{P}) \mathcal{P} (|\varphi_{m_1}\rangle \otimes  \ldots \otimes|\varphi_{m_\eta}\rangle)\, , \quad m_1 < \dots < m_\eta\, ,
    \label{eq:antisym}
\end{equation}
where $S_{\eta}$ is the symmetric group that consists of all the permutations ($\mathcal{P}$) for $\eta$ distinct objects. Preparing such anti-symmetrized states is an initialization step of a first-quantized quantum simulation, with unitary implementations taking the form in Figure \ref{fig:1stquant}. Anti-symmetrization algorithms based on sorting were presented in the original work on the first-quantized simulation\cite{abrams_simulation_1997}, and have been subsequently improved. The asymptotically best-known algorithm for preparing anti-symmetrized states (including superpositions) has gate complexity $\mathcal{O}(\eta \log \eta \log M)$\cite{berry_improved_2018}. 

With the representation of the fermionic states fixed, the next task is to represent the relevant operators. In the context of electronic structure, the real space operators appearing in the Hamiltonian are the kinetic and potential energy operators, e.g., Eq.~(\ref{eq:Helec}). The block-encoding access model is most commonly used to implement these operators on the encoded Hilbert space. One decomposes the operators into the LCU form, and derives circuits that implement each of the unitary terms as in Eq. \ref{eq:prepare}. In fact, much of the cost of a first-quantized simulation is determined by specific block encoding techniques. A common strategy is to choose a basis set that facilitates efficient block encoding\cite{babbushQuantumSimulationChemistry2019, su2021fault}. 

First-quantized simulations are generally not considered amenable to near-term experiments. The antisymmetrization procedure typically does not possess a local nor shallow circuit structure, making it hard to implement on noisy quantum devices. However, algorithm development and resource estimations for first-quantized simulations in the context of the fault-tolerant regime have been an active field. In fact, regarding the resource costing model based on the non-Clifford gate counts, first-quantized simulation algorithms have competitive scaling both for eigenstate problems and Hamiltonian simulations~\cite{babbushQuantumSimulationChemistry2019, su_fault-tolerant_2021,babbush_mean-field_2023,berry2024quantum}. The typical strategy behind these algorithms is to use carefully designed basis sets to discretize the first-quantized Hamiltonian, then perform block encoding methods on the discretized Hamiltonian as in Section~\ref{sec:algorithm}.

\subsection{Second quantization}\label{sec:secondQ}

\begin{figure}[t]
    \centering
     \includegraphics[width=0.65\linewidth]{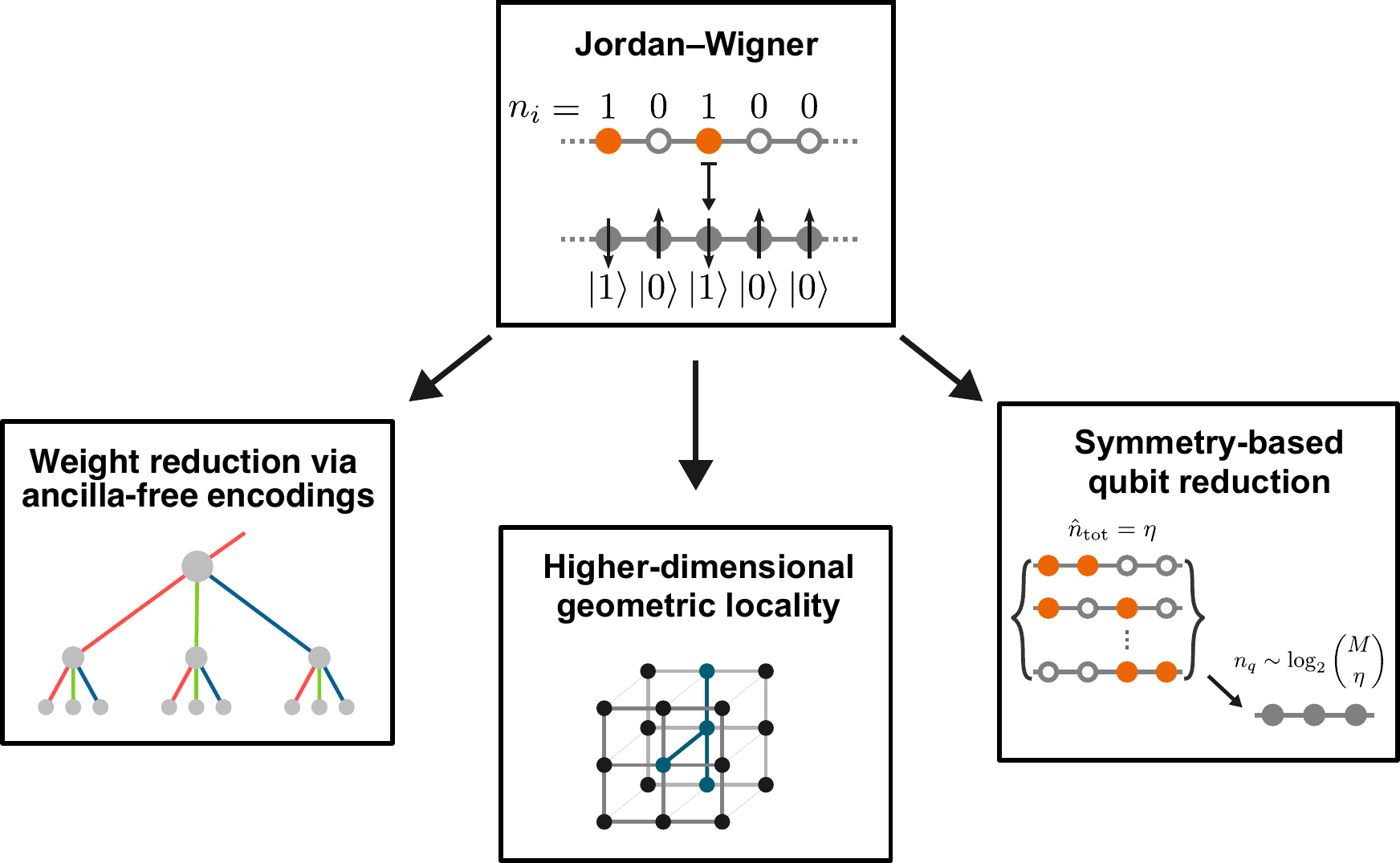}
    \caption{Second-quantized encoding methods.}
    \label{fig:2nd_quantization_guide}
\end{figure}

In second quantization, the basic degrees of freedom are the modes, which in fermionic systems can be either occupied or unoccupied. The $n$--mode Hilbert space, then, has dimension $2^n$, the same as a system of $n$ qubits, with a constraint on physical states known as fermion parity superselection forbidding coherent superpositions of even and odd parity states~\cite{Wick1997}. The algebra of operators on $n$ modes is generated by annihilation (creation) operators $a_i$ ($a_i^{\dagger}$) for $i=1,2,\dots, n$, which add and subtract one from the occupancy of the $i$th mode and whose anticommutation relations are
\begin{align} \label{eqn:cars1}
    \{a_i, a_j\} = 
    a_i a_j + a_j a_i = 
    0 \, , \quad  \{a_i^{\dag}, a_j^{\dag}\} 
    = 0 \, , \quad  \{a_i,a_j^{\dag}\} = \delta_{ij}I\, .
\end{align}
Starting from the vacuum state $\ketf{\vec{0}}$, which obeys $a_i \ketf{\vec{0}} = 0$ and is thus annihilated by all the annihilation operators, the occupation number basis of states corresponding to $\vec{n}\in\mathbb{F}_2^n$ is constructed as
\begin{align}
\ketf{\vec{n}} = (a_{1}^\dagger)^{n_1}(a_{2}^\dagger)^{n_2}\dots (a_{n}^\dagger)^{n_{n}} \ketf{\vec{0}}\, .
\end{align}
Another useful generating set of operators are the $2n$ Majorana operators, defined as
\begin{align}
    \gamma_i = a_i + a_i^{\dag} \, , \quad \overline{\gamma}_i = -\mathrm{i}(a_i - a_i^{\dag})\, .
\end{align}
Like Pauli operators, the Majorana operators are both Hermitian and unitary. They obey the following anticommutation relations
\begin{align}
\{\gamma_i,\gamma_j\} = \{\overline{\gamma}_i,\overline{\gamma}_j\} = 2\delta_{ij}I\, ,  \quad \{\gamma_i,\overline{\gamma}_j\}=0\, . \label{eqn:cars2}
\end{align}
An important element of the above algebra is the total parity operator $(-1)^{\sum_i a_i^{\dag}a_i} = \prod_i (-\mathrm{i}\g_i\gb_i$), which grades the basis states $\ketf{\vec{n}}$ into even (odd) parity, $(\pm 1)$-eigenstates. Products of even (odd) numbers of the Majorana operators (anti)commute with $\prod_i (-\mathrm{i}\g_i\gb_i)$, and thus a notion of parity for the generators of the fermionic algebra also arises from the total parity operator.

To encode a second-quantized fermionic system, we specify a generating set of fermionic operators, $F=\{f_j\}_{j}$, for which we will take products of Majorana operators (due to their similarity to Pauli operators). We then obtain Pauli operators $\widetilde{F}=\{\widetilde{f}_j\}_{j}$ which need only have the same (group) commutation relations. After writing the Hamiltonian (or any operator) as a polynomial in the elements of $F$, we perform the mapping by replacing each $f_j$ with its encoded partner
\begin{align}
    \text{\textbf{Fermion-to-qubit mapping:}} \quad \fh = \sum_{ij}h_{ij} a_i^{\dag} a_j + \sum_{ijkl}h_{ijkl} a_i^{\dag} a_j^{\dag} a_k a_l + \ldots  = \sum_{J} h_J \prod_{j\in J} f_j \quad \longmapsto \quad \widetilde{H} = \sum_{J} h_J \prod_{j\in J} \widetilde{f}_j
    \label{eqn:qubitham}
\end{align}
where we used $J$ as a multi-index. The precise algebraic structure underlying such encodings was described in Refs.\ \citenum{cobanera2011bond} and \citenum{chien2022optimizing}. The simple conditions given above are sufficient to ensure that the mapping constitutes a faithful representation of the $C^*$-algebra of operators generated by $F$.  

Note that the set of operators $F$ need not generate the entire algebra of fermionic operators. It can also be overcomplete. Common choices for the generating set include \textit{(a)} all of the $2n$ Majorana operators $F=\{\g_i,\gb_i\}_{i=1}^{n}$, allowing us to encode the entire fermionic algebra of operators, as in the Jordan--Wigner and Bravyi--Kitaev transformations \cite{jordan1928about, bravyi2002fermionic} as well as other ancilla-free encodings; and \textit{(b)} even-parity products of Majorana operators corresponding to couplings on a graph $G=(V,E)$ with local parity operators for the modes at the vertices and nearest--neighbor couplings on the edges $F =\{ -\mathrm{i}\g_i\gb_i\} _{i\in V} \bigcup \{-\mathrm{i}\g_j\g_k\}_{(j,k)\in E}$. When only even--parity operators are included in $F$, we may not have straightforward access to the odd--parity fermionic operators, but this is not an issue as all observables have even parity.

Expressing qubit Hamiltonians in the form of Eq.\ (\ref{eqn:qubitham}) is useful for product formula time evolution as Pauli operators can be exponentiated simply. It is also useful in the block encoding framework as the Hamiltonian is already in the linear combination of unitaries (LCU) form and we can utilize the $\prep$-$\sel$ oracles in Eq.\ (\ref{eq:prepare}) to block-encode the Hamiltonian.

Just as the fermionic vacuum is the state stabilized by the local parity operators, the encoded vacuum state is the state stabilized by the encoded local parity operators. This can be taken as the starting point in the correspondence between states, with the rest of the states constructed by acting with encoded operators. The encoded system may also have more qubits than the fermionic modes, in which case we must restrict to a subspace stabilized by Pauli operators that act as the identity on the encoded system. 

Up to parity constraints, any two second-quantized fermion-to-spin encodings that map Majoranas to Paulis are equivalent via a Clifford circuit and the addition or removal of ancilla qubits. If the cost measure of a computation counts only logical non-Clifford gates, then the collection of all second-quantized mappings is essentially equivalent, up to differences in the number of logical qubits. On the other hand, cost models that count logical Clifford gates or value parallelizability will favor certain second-quantized mappings over others and much of the literature is devoted to devising encodings that reduce the Pauli weight of encoded observables.

In the rest of this subsection, we review the most widely--used correspondence between fermionic and spin systems, the Jordan--Wigner transformation, before turning to other mappings that require the same number of qubits. Finally, we discuss mappings that utilize additional qubits in order to preserve higher--dimensional spatial locality structures.

\subsubsection{The Jordan--Wigner transformation} \label{sec:jw}

The most familiar mapping of fermionic systems to spin systems is the Jordan--Wigner transformation\cite{jordan1928about}. We can define it as a map of all the Majorana operators for an $n$-mode fermionic system to the Pauli operators for an $n$-qubit system:
\begin{align}
    \substack{\text{\textbf{Jordan--Wigner}} \\ \text{\textbf{ transformation:}}} \quad \ketf{\vec{n}} \longmapsto \ket{\vec{n}} \quad \implies \quad 
    \begin{cases}\gamma_i \longmapsto \left(\prod_{j<i}Z_j \right)X_i\, , \\
    \overline{\gamma}_i  \longmapsto \left( \prod_{j<i}Z_j \right) Y_i
    \end{cases} \implies \quad \gamma_i \gamma_j \longmapsto \mathrm{i} \, Y_j \left(\prod_{j<k<i} Z_k\right) X_i\, ,
    \label{eqn:jw}
\end{align}
where the expression $\ket{\vec{n}}$ denotes the computational basis state with $n$-bit string $\vec{n}$.
The mapping in Eq.\ (\ref{eqn:jw}) preserves the commutation relations between the fermionic operators and their encoded counterparts due to the $Z$-strings. The Jordan--Wigner transformation is well-suited to fermionic Hamiltonians that are local in one dimension, which consist only of interactions between modes with consecutive labels $i$ and $i+1$. Such Hamiltonians transform into local spin chain Hamiltonians under the Jordan--Wigner transformation as the $Z$-strings cancel for each encoded term. Note that mapping constant-weight fermionic operators to constant-weight qubit operators with no space overhead is generally not possible beyond special cases such as chains and loop-free bounded-degree graphs.

We can of course use the Jordan--Wigner transformation for any fermionic system, regardless of the geometric structure of its interactions or its dimensionality by simply enumerating the modes and treating the system as if it were a 1D fermionic system with long-range interactions. The long-range fermionic operators, involving interactions between distant modes $i$ and $j$, will be mapped to nonlocal qubit operators as in Eq.\ (\ref{eqn:jw}). For a generic fermionic Hamiltonian, the resulting encoded terms will act on $\mathcal{O}(n)$ qubits in general. Such high-weight operators will consume $\mathcal{O}(n)$ two-qubit (typically Clifford) gates in a product formula time evolution or in the construction of a $\sel$ oracle as in Eq.\ (\ref{eq:prepare}). 

There are a number of strategies available for mitigating the non-locality of the Jordan--Wigner transformation. First, since the linear ordering of the fermionic modes is arbitrary, we are free to relabel the modes as we like before applying the Jordan--Wigner transformation. By making a more judicious choice of ordering, we can often reduce the total weight of the Pauli terms in an encoded Hamiltonian\cite{chiew2023discovering, parelladilme2024reducing}. Second, non-locality can be addressed during circuit compilation. Fermionic swap networks are one strategy for reducing a circuit involving long--range terms on a fixed ordering to one involving only short-range terms on a chain with a dynamic ordering. The method employs fermionic swap gates $\fswap = \cz\cdot\swap$ to exchange the positions of interacting modes until they are nearby on the chain. This strategy typically requires cycling through a number of orderings on the order of the longest-range term in the Hamiltonian\cite{cade2020fermihubbard}. In the LCU framework, the polynomial circuit depth from the high-weight terms in the qubit Hamiltonian can also be mitigated
\cite{babbush2018encoding,wan2021exponentially}.

\subsubsection{Ancilla-free encodings}\label{sec:Q2-1to1}

Using mappings between the $n$-mode and $n$-qubit spaces that yield lower-weight representations of the fermionic operators than the Jordan--Wigner transformation may simplify circuit compilation, especially for fermionic systems that do not exhibit one-dimensional interaction structures. These \textit{ancilla-free encodings} are particularly useful for systems when there is no spatial locality structure manifest in the second-quantized Hamiltonian, or when the number of available logical qubits is limited. Early examples of these alternatives include the Bravyi--Kitaev\cite{bravyi2002fermionic} and parity basis \cite{Seeley_2012} transformations. With very few exceptions\cite{steudtner2018fermion}, almost all second-quantized encodings in the literature represent Majorana operators as Pauli operators. 

A fact that can sometimes be useful is that the transformed operators of these encodings take the form $\gamma_i \mapsto U \cdot \left( \prod_{j<i} Z_j \right)  X_i \cdot U^\dagger$ and $\overline{\gamma}_i \mapsto U \cdot \left( \prod_{j<i} Z_j \right) Y_i \cdot U^\dagger$ for an $n$-qubit unitary $U$. Restricting this unitary to $\cnot$ circuits yields a subclass of ancilla-free encodings referred to as \textit{linear} encodings, which have the property of transforming the occupation number basis as $\ketf{\vec{n}} \mapsto \ket{G \cdot \vec{n}}$, for $G\in \mathrm{GL}(n,\mathbb{F}_2)$\cite{bravyi2017tapering, steudtner2018fermion, Wang2021resourceoptimized,wang2023evermore}. Consequently, for linear encodings, the fermionic vacuum state corresponds to the all-zero computational basis state, $\ketf{\vec{0}}\mapsto|00\ldots 0\rangle$, which may reduce state initialization costs\cite{miller2023bonsai, miller2024treespilation}. 

\begin{figure}
    \centering
    \includegraphics[width=0.9\linewidth]{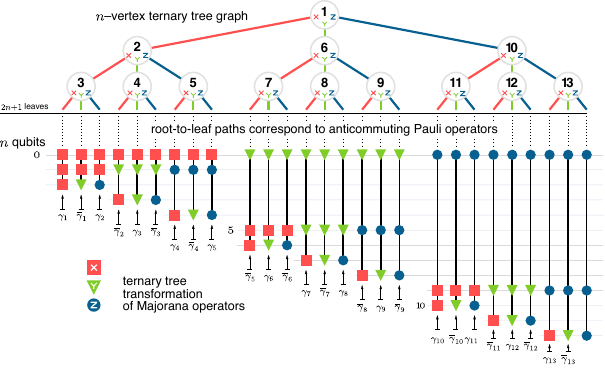}
    \caption{The ternary tree transformation  
    $\gamma_i, \overline{\gamma}_i \mapsto \pm P_i, \overline{P}_i$ is an ancilla-free encoding that maps from Majorana to Pauli operators using the structure of a tree graph.
    }
    \label{fig:lineartree}
\end{figure}

Within the class of ancilla-free second-quantized encodings, we highlight the ternary tree transformation for it simplicity, ability to capture the optimal average Pauli weight for encoded Majorana operators, and readiness with which it can be optimized\cite{jiang2020optimal,Vlasov_2022}. To define this mapping, add $2n{+}1$ leaves to an $n$--vertex ternary tree graph and associate a qubit to each vertex, as in Figure \ref{fig:lineartree}.
The paths from the root vertex to the leaves can diverge in three directions at each vertex, corresponding to the three anticommuting Pauli operators. The encoded Majorana operators are constructed from the product of the Paulis on each qubit in the path taken from root to leaf, with weights upper--bounded by the depth of the tree. Any ternary tree graph and any assignment of Majoranas to the Pauli strings provides a valid mapping, and can be made into a linear encoding \cite{chiew2024ternarytreetransformationsequivalent}. If the ternary tree is complete, as in Figure \ref{fig:lineartree}, the encoded Majoranas each have a Pauli weight of $\lceil\log_3(2n+1)\rceil$ qubits, which is the provably minimal average weight. 

Further optimizations of ancilla-free encodings have been proposed. Data structures such as Fenwick and Sierpinski trees have also provided an inspiration for linear encodings\cite{havlivcek2017operator, harrison2024sierpinskitrianglefermiontoqubittransform}, corresponding to different choices for the invertible binary matrix $G$ that yield Majorana operators of lower Pauli weight. There has been algorithmic optimization on the ternary tree graph structure itself to reduce $\swap$\cite{miller2023bonsai} and $\cnot$ gate counts \cite{miller2024treespilation} in ancilla-free fermionic simulation \cite{miller2023bonsai, miller2024treespilation}. Optimizing over Clifford subsets for the unitary $U$ is another strategy that is not limited to the linear encoding family\cite{yu2025cliffordcircuitbasedheuristic}. Another considered strategy is to try to choose a mapping that minimizes the entanglement in the encoded state, which may reduce the cost of certain state preparation techniques. \cite{parelladilme2024reducing,miller2024treespilation}.

\subsubsection{Symmetry-based qubit reduction}\label{sec:fixed_particle_number}

In the presence of a symmetry, ergodicity is broken and dynamics can be restricted to subspaces of smaller dimension than the total Hilbert space. That is, the Hilbert space decomposes into a direct sum of charge sectors that do not mix. This can be leveraged to reduce the number of qubits used in a simulation by encoding only the relevant charge sectors. Intuitively, the larger the symmetry group acting on the system, the more the dynamics can be restricted and the fewer qubits that can potentially be required for simulating the symmetric dynamics.

In the simplest case, the symmetry is simply a $\mathbb{Z}_2$ symmetry acting as a tensor product of Pauli operators. The Hilbert space splits into even and odd subspaces of equal dimension. By a Clifford unitary transformation, one of the qubits of the system can be made to specify the $\mathbb{Z}_2$ charge of the system, while all symmetric operators act at most diagonally on the charge qubit. Qubit tapering is a method for removing a single qubit for each independent $\mathbb{Z}_2$ factor in the symmetry group\cite{bravyi2017tapering}. A common example that arises is the total fermion parity, which after a Jordan--Wigner transformation is $P_{tot} = Z_j^{\otimes}$. In this case, a single qubit can be tapered to encode only the even parity sector. Other examples include point-group symmetries which are generally non-abelian and a maximal abelian subgroup is used\cite{setia2020reducing}.

Another common example is $U(1)$ particle number symmetry. In this case, the total $M$-mode Hilbert space decomposes into a direct sum of subspaces with fixed numbers of particles $\eta$, each of dimension $(\substack{M\\\eta})$. For the case of $\eta \ll M$, this can be taken advantage of to dramatically reduce the qubit requirements from $M$ to $\mathcal{O}(\log M)$ -- reminiscent of first quantization where the dependence on the number of modes is similarly mild. This symmetry has been exploited to decrease qubit counts by varying amounts\cite{bravyi2017tapering,steudtner2018fermion,Shee_qubit_efficient,harrison2023reducingqubitrequirementjordanwigner}. It has also been shown that it is possible to achieve qubit requirements polylogarithmic in $(\substack{M\\ \eta})$ as well as polylogarithmic gate complexity for simulation primitives \cite{kirby_secondquant}. Recently, it was also shown that the minimum qubit requirements can be achieved up to an additive constant while maintaining polylogarithmic gate complexity\cite{carolan2024succinctfermiondatastructures}.

\subsubsection{Local encodings}\label{sec:geometrically_local}

Often, we deal with systems with not just few-body interactions (``$k$-locality''), but interactions that are localized on a lattice or low-degree graph. This geometric locality can be exploited by algorithms to provide asymptotic speedups to tasks such as Hamiltonian evolution, stemming from e.g.\ Lieb-Robinson bounds on information propagation\cite{lieb1972finite,hastings2010locality,haah2021quantum,tran2019locality}. By utilizing a mapping that preserves the \textit{geometric} locality -- mapping a locally-interacting fermionic lattice system to a locally-interacting spin system -- we can directly utilize such algorithms. Additionally, one has access to scheduling speedups, since Hamiltonian terms share support with only few others so that many operations can be executed in parallel\cite{bringewatt2023parallelization}. Trotter steps can in principle be executed in a time independent of the total system size for lattice systems in any fixed dimension. Mappings of this type are called \textit{local encodings}.

Local encodings exist in any number of spatial dimensions, with the Jordan--Wigner transformation being an example in one dimension. Beyond one dimension, they typically use a constant-factor more qubits than fermionic modes, with the overhead depending roughly on the vertex degree of the lattice. The discrepancy in Hilbert space dimension is addressed by the restriction that one is in the $(+1)$-eigenspace of a stabilizer with generators corresponding to loops on the graph. This ensures that, when transporting a particle from one side of the loop to the other, one can take either route with the same result. Local encodings have been optimized to minimize the weight of the encoded operators, to reduce the qubit overhead ratio, for flexibility to be applicable on any interaction graph, and to make symmetries manifest\cite{verstraete2005mapping,steudtner2019quantum,setia2019superfast,jiang2019majorana,chen2019exactthree,chen2020exactarb,chien2020custom,derby2021compact,derby2021compactalt,chien2022optimizing,li2022higher,bringewatt2023parallelization,chen2024error}. The presence of stabilizers makes local encodings into stabilizer codes\cite{gottesman1997stabilizer} and this has been explored to construct higher-distance codes for error-mitigation\cite{setia2019superfast,jiang2019majorana,chen2024error,chien2023simulating,simkovic2024low} and fault-tolerant simulations with logical fermionic modes\cite{landahl2021logical}.

Only even-parity fermionic operators commute with each other when they do not share support, so we should expect that only even-parity operators will be able to be encoded as local spin operators. This is not a severe limitation though as these are the physical observables. A useful (overcomplete) generating set of even-parity operators for fermions on a graph are single-mode parity operators known as vertex operators, $ V_j = -\mathrm{i} \gamma_{j} \overline{\gamma}_j = (-1)^{a_j^\dagger a_j}$,
and nearest-neighbor quadratic couplings called edge operators, $ E_{j,k} =  -\mathrm{i}\gamma_j \gamma_k = -\mathrm{i}(a_j a_k + a_j^\dagger a_k + a_j a_k^{\dag} + a_j^\dagger a_k^\dagger )$. These operators are both Hermitian and unitary and are mapped to Pauli operators. 

\begin{figure}
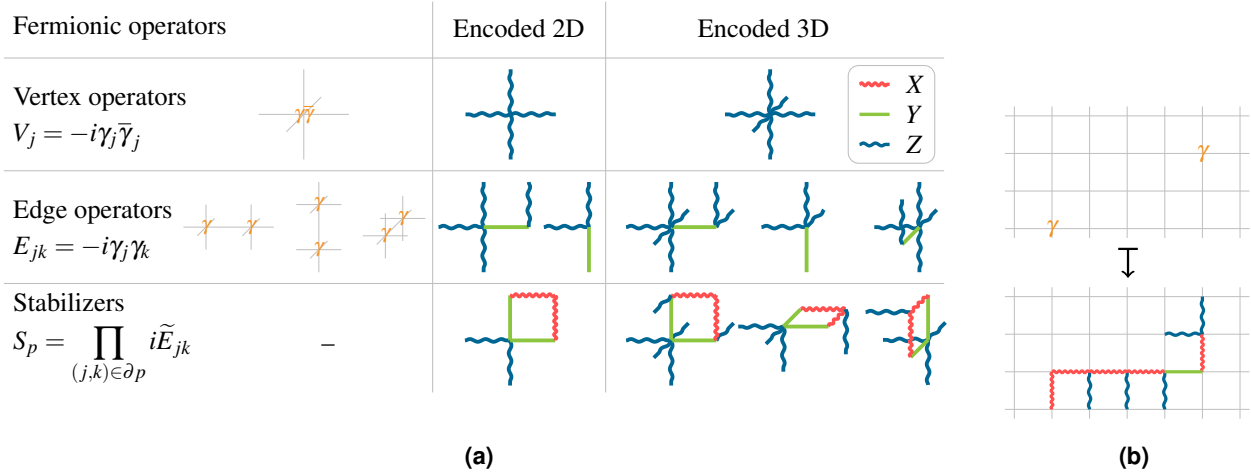

    \centering
    \begin{subfigure}[b]{0.72\textwidth}
        \centering
        \begin{tikzpicture}[line join=round, decoration={snake, amplitude=.18mm, segment length=0.9mm}]

\draw (-9.5,1.15) node[anchor=west] {Fermionic operators};
\draw (-3.65,1.15) node[anchor=west] {Encoded 2D};
\draw (-0.4,1.15) node[anchor=west] {Encoded 3D};

\node at (2.3,0) {\input{figures/pauli_legend}};

\draw (-9.5,0.2) node[anchor=west] {Vertex operators};
\draw (-9.5,-0.3) node[anchor=west] {$V_j = -i \gamma_j \overline{\gamma}_j$};
\draw (-9.5,-1.3) node[anchor=west] {Edge operators};
\draw (-9.5,-1.8) node[anchor=west] {$E_{jk} = -i \gamma_j \gamma_k$};
\draw (-9.5,-2.5) node[anchor=west] {Stabilizers};
\draw (-9.5,-3.2) node[anchor=west] {$\displaystyle S_p = \prod_{(j,k)\in \partial p} i\widetilde{E}_{jk}$};

\draw (-5.4,-3.1) node[anchor=west] {--};
\node at (-5.5,0) {\scalebox{0.3}{\input{figures/vertex_maj}}};
\node at (-6.5,-1.5) {\scalebox{0.3}{\input{figures/E_x_maj}}};
\node at (-5.3,-1.5) {\scalebox{0.3}{\input{figures/E_y_maj}}};
\node at (-4.3,-1.5) {\scalebox{0.3}{\input{figures/E_z_maj}}};

\node at (-2.75,0) {\scalebox{0.3}{\input{figures/2d_f_vertex}}};
\node at (-3.1,-1.5) {\scalebox{0.3}{\input{figures/2d_f_string_gg_x}}};
\node at (-2,-1.5) {\scalebox{0.3}{\input{figures/2d_f_string_gg_y}}};
\node at (-2.75,-3) {\scalebox{0.3}{\input{figures/2d_f_stab}}};

\node at (0.7,0) {\scalebox{0.3}{\input{figures/3d_f_vertex}}};
\node at (-0.5,-1.5) {\scalebox{0.3}{\input{figures/3d_f_string_gg_x}}};
\node at (1,-1.5) {\scalebox{0.3}{\input{figures/3d_f_string_gg_y}}};
\node at (2.5,-1.5) {\scalebox{0.3}{\input{figures/3d_f_string_gg_z}}};
\node at (-0.5,-3) {\scalebox{0.3}{\input{figures/3d_f_stab_xy}}};
\node at (1,-3) {\scalebox{0.3}{\input{figures/3d_f_stab_xz}}};
\node at (2.5,-3) {\scalebox{0.3}{\input{figures/3d_f_stab_yz}}};

\draw[color=lightgray] (-3.8,1.5) -- (-3.8,-3.7);

\draw[color=lightgray] (-1.5,1.5) -- (-1.5,-3.7);

\draw[color=lightgray] (-9.5,0.75)--(3,0.75);
\draw[color=lightgray] (-9.5,-0.75)--(3,-0.75);
\draw[color=lightgray] (-9.5,-2.25)--(3,-2.25);

\end{tikzpicture}
        \label{subfig:2d3d_ops}
        \caption{}
    \end{subfigure}
    \begin{subfigure}[b]{0.26\textwidth}
        \centering
        \begin{tikzpicture}[line join=round, decoration={snake, amplitude=.6mm, segment length=3mm}]

    \node at (0,0) {\scalebox{0.25}{\begin{tikzpicture}[line join=round, decoration={snake, amplitude=.6mm, segment length=3mm}]

    \draw[color=lightgray,line width=0.4mm] (-0.5,0) -- (12.5,0);
    \draw[color=lightgray,line width=0.4mm] (-0.5,2) -- (12.5,2);
    \draw[color=lightgray,line width=0.4mm] (-0.5,4) -- (12.5,4);
    \draw[color=lightgray,line width=0.4mm] (-0.5,6) -- (12.5,6);

    \draw[color=lightgray,line width=0.4mm] (0,-0.5) -- (0,6.5);
    \draw[color=lightgray,line width=0.4mm] (2,-0.5) -- (2,6.5);
    \draw[color=lightgray,line width=0.4mm] (4,-0.5) -- (4,6.5);
    \draw[color=lightgray,line width=0.4mm] (6,-0.5) -- (6,6.5);
    \draw[color=lightgray,line width=0.4mm] (8,-0.5) -- (8,6.5);
    \draw[color=lightgray,line width=0.4mm] (10,-0.5) -- (10,6.5);
    \draw[color=lightgray,line width=0.4mm] (12,-0.5) -- (12,6.5);



    \node at (2,0,0) {\scalebox{3.5}{$\color{BurntOrange}\gamma$}};
    \node at (10,4,0) {\scalebox{3.5}{$\color{BurntOrange}\gamma$}};

\end{tikzpicture}}};

    \draw[|->,line width=0.3mm] (0,-1) -- (0,-1.4);
    
    \node at (0,-2.4) {\scalebox{0.25}{\begin{tikzpicture}[line join=round, decoration={snake, amplitude=.6mm, segment length=3mm}]

    \draw[color=lightgray,line width=0.4mm] (-0.5,0) -- (12.5,0);
    \draw[color=lightgray,line width=0.4mm] (-0.5,2) -- (12.5,2);
    \draw[color=lightgray,line width=0.4mm] (-0.5,4) -- (12.5,4);
    \draw[color=lightgray,line width=0.4mm] (-0.5,6) -- (12.5,6);

    \draw[color=lightgray,line width=0.4mm] (0,-0.5) -- (0,6.5);
    \draw[color=lightgray,line width=0.4mm] (2,-0.5) -- (2,6.5);
    \draw[color=lightgray,line width=0.4mm] (4,-0.5) -- (4,6.5);
    \draw[color=lightgray,line width=0.4mm] (6,-0.5) -- (6,6.5);
    \draw[color=lightgray,line width=0.4mm] (8,-0.5) -- (8,6.5);
    \draw[color=lightgray,line width=0.4mm] (10,-0.5) -- (10,6.5);
    \draw[color=lightgray,line width=0.4mm] (12,-0.5) -- (12,6.5);



    \draw[Z] (4,0) -- (4,2);
    \draw[Z] (6,0) -- (6,2);
    \draw[Z] (8,0) -- (8,2);

    \draw[Z] (8,4) -- (10,4);
    \draw[Z] (10,4) -- (10,6);

    \draw[X] (2,0) -- (2,2);
    \draw[X] (10,2) -- (10,4);

    \draw[X] (2,2) -- (4,2);
    \draw[X] (4,2) -- (6,2);
    \draw[X] (6,2) -- (8,2);
    
    \draw[Y] (8,2) -- (10,2);
\end{tikzpicture}}};

\end{tikzpicture}
        \label{subfig:2d_long_range}
        \caption{}
    \end{subfigure}
    \caption{(a) Generating operators and stabilizers of a locality preserving encoding in 2D and 3D. In the fermionic system, modes are associated to vertices. In this encoding, qubits are on edges, with Pauli operators indicated by highlighted edges.
    (b) Long-range quadratic operators are mapped to deformable string operators with weight linear in the distance between the targeted vertices by taking products of the generating operators.}
    \label{fig:local_ops}
\end{figure}

In Figure \ref{fig:local_ops}(a), we show the encoded generating vertex and edge operators on a square and cubic lattice for an encoding which has appeared a number of times in the literature under the names Bravyi--Kitaev superfast encoding and exact bosonization\cite{bravyi2002fermionic,levin2003fermions,havlivcek2017operator,chen2018exacttwo}. The fermionic modes are associated to vertices of a lattice and the qubits are associated to edges. The encoded edge operators that transport particles and create them in pairs are interpreted as short strings of Pauli $X$ operators between the targeted vertices which are dressed by $Z$s to produce the correct commutation relations. Long-range operators become longer string operators on the lattice as in Figure \ref{fig:local_ops}.(b) which can be freely deformed by multiplying with stabilizers. This is the higher-dimensional version of the Jordan--Wigner string. 

The string-like nature hints at connections with topological order, and indeed there are close connections between local encodings and the toric code\cite{kitaev2003fault}. Tools for studying topological order and quantum codes have also found use in constructing local encodings\cite{haah2016algebraic,chen2023equivalence,chien2022optimizing,chen2024error,simkovic2024low}. A number of encodings have been shown explicitly to be related by finite-depth Clifford circuits\cite{chen2023equivalence}. It has also been shown that long-range entanglement is necessary for a local encoding of fermions\cite{guaita2024locality}. Consequently, unitary initialization of the vacuum state requires $\mathcal{O}(L)$ time for an $L\times L$ lattice or $\mathcal{O}(1)$ time with measurements and feedback via error correction\cite{bravyi2006lieb}.

The general structure illustrated here is shared amongst all local encodings, although the details differ, leading to different costs in terms of qubits and operator weights. Different mappings place qubits on different parts of the lattice (vertices, edges, faces, etc.) and the generating operators are constructed differently. The compact encodings \cite{derby2021compact,derby2021compactalt} are particularly efficient in terms of qubit/mode ratio and operator weight for square and cubic lattices. For general graphs the generalized superfast encoding \cite{setia2019superfast} provides the lowest encoded operator weights of any known mapping. Sometimes, strict operator locality can be relaxed when an algorithm requires that encoded operators be local only on some scale between the total system size and the lattice scale, and concatenation-like schemes can be devised to reduce qubit overhead \cite{steudtner2018fermion,haah2021quantum,chien2020custom,clinton2024towards,rhodes2024exponential, o2024ultrafast}. 

\subsection{Comparison of encoding methods}\label{sec:comparison}

In this section, we introduced encoding schemes based on both first and second quantization. The different encoding methods can impose different preferences and cost scaling, and deciding which one is preferable is part of the holistic algorithm design process. In this final subsection, we summarize the various factors that should be taken into account. For comparing first- and second-quantized representations, we focus specifically on systems to which either formalism can be applied, such as the electronic Hamiltonian. We also compare aspects of the different second-quantized encoding methods.

In quantum chemistry, one typically considers the particle number to be fixed, but is interested in systematically minimizing the Hamiltonian discretization error with respect to basis set size. Consequently, lower resource scaling in terms of the number of basis functions is crucial. Both the first-quantized representation (Sec.\ \ref{sec:firstQ}) as well as recent, specialized second-quantized encodings (Sec.\ \ref{sec:fixed_particle_number}) are able to encode fermionic states of fixed particle number with qubit counts scaling poly-logarithmically in the basis set size. One is also able to implement standard algorithmic primitives, such as Hamiltonian simulation, with gate complexity sublinear in the basis set size in these methods. Therefore, they can offer a significant advantage over standard second-quantized encodings for such systems, both in terms of qubit count and gate complexity.
However, algorithms formulated in first quantization have been subjected to more comprehensive resource estimation studies than the fixed particle number second-quantized encodings\cite{babbush_exponentially_2018, babbushQuantumSimulationChemistry2019, su_fault-tolerant_2021}. 

For systems most naturally formulated in second quantization, or when particle number is not even conserved, one still has choices in how to represent the system. A second-quantized encoding which preserves the spatial locality of the system (Sec.\ \ref{sec:geometrically_local}) in question can be used to implement Trotter steps or perform measurements in constant-depth, particularly when the interactions of the encoded Hamiltonian match the geometry of the quantum computing architecture. For second-quantized systems without a spatial structure or when one cannot afford the constant-factor qubit overhead of local encodings, ancilla-free encodings, such as the ternary tree transformation introduced in Sec.\ \ref{sec:Q2-1to1}, can be used to provide a low-weight (typically logarithmic in system size) representation of the encoded Hamiltonian. For algorithms utilizing the Hamiltonian expressed as a sum of Paulis, different second-quantized methods will use the same number of non-Clifford gates but can use a number of non-Clifford gates which differs by a polynomial factor in the system size.

\section{Outlook}

At present, quantum computers are not yet practical tools for mainstream calculations in chemistry and physics.  Sufficiently accurate computations of relevant observables of chemical systems and materials of industrial interest will likely require $\sim 10^3$ logical qubits and current gate estimates are at least $\sim10^8$ using currently known methods\cite{low2025fast}. This puts these applications well into the regime of large-scale fault-tolerant quantum computation. Current state of the art estimates for lattice gauge theories of interest in nuclear physics are higher still, with logical qubit counts up to $\sim10^{11}$ and up to $\sim10^{21}$ gates\cite{rhodes2024exponential}. 
Work bringing these numbers down is ongoing, and we can hope for significant improvements.

Model condensed matter systems such as the two-dimensional Hubbard model are thought to be a much closer target for quantum simulation due to their simple structure. While a fault-tolerant quantum computer is still likely required in order to be competitive with existing classical methods, there have been a number of experiments with noisy devices simulating the dynamics of these model systems and attempting to prepare low-energy states via variational means. For simple model systems in one or two dimensions, near-term simulations are helped by the fact that the device coupling graphs (which are often two-dimensional) contain the interaction graph of the encoded Hamiltonians. In recent demonstrations in noisy experiments, utilizing a local encoding led to improved calculations of energy density~\cite{nigmatullin2024experimental} and more scalable studies of Floquet dynamics in two dimensions~\cite{evered2025probing}.

There are a number of interesting alternative approaches that we did not discuss in this review (e.g. analog simulators~\cite{tan2022analog}). Certain platforms, such as trapped ions, allow for the direct manipulation of fermionic degrees of freedom. These can be used to emulate the physics of systems of interest via Hamiltonian and Floquet engineering without having to construct the dynamics from discrete gates. As the degree of control in these systems becomes greater, the divide between analog and digital devices shrinks, allowing for novel hybrid analog-digital modes of operation~\cite{liu2024toward,crane2024hybrid}. Such methods may play a greater role in scientific inquiry enabling us to study model systems such as the Hubbard model, even without quantum error correction. 

Efficient methods of encoding the degrees of freedom onto quantum computers will undoubtedly play a role in further developments reducing the quantum resources required for scientifically and industrially relevant computations. While we have focused on fermionic degrees of freedom in this review, improved methods of encoding and performing computations on other degrees of freedom will also have benefits across physics.

\section*{Acknowledgments}
R.W.C. acknowledges support by U.S. Department of Energy, Office of Science, National Quantum Information Science Research Centers, Quantum Systems Accelerator and U.S. Department of Energy, Office of Science, Office of Advanced Scientific Computing Research, Exploratory Research for Extreme Scale Science program. Sandia National Laboratories is a multi-mission laboratory managed and operated by National Technology \& Engineering Solutions of Sandia, LLC (NTESS), a wholly owned subsidiary of Honeywell International Inc., for the U.S. Department of Energy’s National Nuclear Security Administration (DOE/NNSA) under contract DE-NA0003525. This written work is authored by an employee of NTESS. The employee, not NTESS, owns the right, title and interest in and to the written work and is responsible for its contents. Any subjective views or opinions that might be expressed in the written work do not necessarily represent the views of the U.S. Government. The publisher acknowledges that the U.S. Government retains a non-exclusive, paid-up, irrevocable, world-wide license to publish or reproduce the published form of this written work or allow others to do so, for U.S. Government purposes. The DOE will provide public access to results of federally sponsored research in accordance with the DOE Public Access Plan.
S.S. was supported by the Royal Society University Research Fellowship and ``Quantum simulation algorithms for quantum chromodynamics'' grant (ST/W006251/1). 
M.M. acknowledges funding by RQS QLCI grant OMA-2120757. 
M.L.C. received the support of a Cambridge Australia Allen \& DAMTP Scholarship during the preparation of this document.
C.M. acknowledges support from the Intelligence Advanced Research Projects Activity (IARPA), under the Entangled Logical Qubits program through Cooperative Agreement Number W911NF-23-2-0223. The views and conclusions contained in this document are those of the authors and should not be interpreted as representing the official policies, either expressed or implied, of IARPA, the Army Research Office, or the U.S. Government. The U.S. Government is authorized to reproduce and distribute reprints for Government purposes notwithstanding any copyright notation herein.
J.D.W., J.N.,  B.H.,  W.W., T.M.H. and G.E.S. acknowledge support by the U.S. Department of Energy, Office of Basic Energy Sciences, under Award DE-SC0019374.  G.E.S. is a Welch Foundation Chair (C-0036). W.W. and J.D.W. were also partially supported by ARO grant W911NF2410043.
J.D.W. holds concurrent appointments at Dartmouth College and as an Amazon Visiting
 Academic. This paper describes work performed at Dartmouth College and is not associated with Amazon.

\bibliography{main}

@article{bosse2025efficient,
  title={Efficient and practical Hamiltonian simulation from time-dependent product formulas},
  author={Bosse, Jan Lukas and Childs, Andrew M. and Derby, Charles and Gambetta, Filippo Maria and Montanaro, Ashley and Santos, Raul A.},
  journal={Nature Communications},
  volume={16},
  number={1},
  pages={2673},
  year={2025},
  publisher={Nature Publishing Group},
  doi={10.1038/s41467-025-57580-5},
  url={https://www.nature.com/articles/s41467-025-57580-5}
}

@article{levin2003fermions,
  title={Fermions, strings, and gauge fields in lattice spin models},
  author={Levin, Michael and Wen, Xiao-Gang},
  journal={Physical Review B},
  volume={67},
  number={24},
  pages={245316},
  year={2003},
  publisher={APS}
}

@article{kitaev2003fault,
  title={Fault-tolerant quantum computation by anyons},
  author={Kitaev, A Yu},
  journal={Annals of physics},
  volume={303},
  number={1},
  pages={2--30},
  year={2003},
  publisher={Elsevier}
}

@article{bravyi2002fermionic,
  title={Fermionic quantum computation},
  author={Bravyi, Sergey B and Kitaev, Alexei Yu},
  journal={Annals of Physics},
  volume={298},
  number={1},
  pages={210--226},
  year={2002},
  publisher={Elsevier}
}

@article{setia2019superfast,
  title={Superfast encodings for fermionic quantum simulation},
  author={Setia, Kanav and Bravyi, Sergey and Mezzacapo, Antonio and Whitfield, James D},
  journal={Physical Review Research},
  volume={1},
  number={3},
  pages={033033},
  year={2019},
  publisher={APS}
}

@article{hastings2010locality,
  title={Locality in quantum systems},
  author={Hastings, Matthew B},
  journal={Quantum Theory from Small to Large Scales},
  volume={95},
  pages={171--212},
  year={2010}
}

@article{chen2020exactarb,
  title={Exact bosonization in arbitrary dimensions},
  author={Chen, Yu-An},
  journal={Physical Review Research},
  volume={2},
  number={3},
  pages={033527},
  year={2020},
  publisher={APS}
}

@article{chen2018exacttwo,
  title={Exact bosonization in two spatial dimensions and a new class of lattice gauge theories},
  author={Chen, Yu-An and Kapustin, Anton and Radi{\v{c}}evi{\'c}, {\DJ}or{\dj}e},
  journal={Annals of Physics},
  volume={393},
  pages={234--253},
  year={2018},
  publisher={Elsevier}
}

@article{chen2019exactthree,
  title={Bosonization in three spatial dimensions and a 2-form gauge theory},
  author={Chen, Yu-An and Kapustin, Anton},
  journal={Physical Review B},
  volume={100},
  number={24},
  pages={245127},
  year={2019},
  publisher={APS}
}

@article{verstraete2005mapping,
  title={Mapping local Hamiltonians of fermions to local Hamiltonians of spins},
  author={Verstaete,F. and Cirac, J. I.},
  journal={Journal of Statistical Mechanics: Theory and Experiment},
  volume={2005},
  number={09},
  pages={P09012},
  year={2005},
  publisher={IOP Publishing}
}

@article{steudtner2018fermion,
  title={Fermion-to-qubit mappings with varying resource requirements for quantum simulation},
  author={Steudtner, Mark and Wehner, Stephanie},
  journal={New Journal of Physics},
  volume={20},
  number={6},
  pages={063010},
  year={2018},
  publisher={IOP Publishing}
}

@article{chen2023equivalence,
  title={Equivalence between fermion-to-qubit mappings in two spatial dimensions},
  author={Chen, Yu-An and Xu, Yijia and others},
  journal={PRX Quantum},
  volume={4},
  number={1},
  pages={010326},
  year={2023},
  publisher={APS}
}

@article{derby2021compact,
  title={Compact fermion to qubit mappings},
  author={Derby, Charles and Klassen, Joel and Bausch, Johannes and Cubitt, Toby},
  journal={Physical Review B},
  volume={104},
  number={3},
  pages={035118},
  year={2021},
  publisher={APS}
}

@article{derby2021compactalt,
  title={A compact fermion to qubit mapping part 2: {A}lternative lattice geometries},
  author={Derby, Charles and Klassen, Joel},
  journal={arXiv preprint arXiv:2101.10735},
  year={2021}
}

@article{chien2022optimizing,
  title={Optimizing fermionic encodings for both hamiltonian and hardware},
  author={Chien, Riley W and Klassen, Joel},
  journal={arXiv preprint arXiv:2210.05652},
  year={2022}
}

@article{guaita2024locality,
  title={On the locality of qubit encodings of local fermionic modes},
  author={Guaita, Tommaso},
  journal={arXiv preprint arXiv:2401.10077},
  year={2024}
}

@article{chien2020custom,
  title={Custom fermionic codes for quantum simulation},
  author={Chien, Riley W and Whitfield, James D},
  journal={arXiv preprint arXiv:2009.11860},
  year={2020}
}

@article{chiew2023discovering,
  title={Discovering optimal fermion--qubit mappings through algorithmic enumeration},
  author={Chiew, Mitchell and Strelchuk, Sergii},
  journal={Quantum},
  volume={7},
  pages={1145},
  year={2023},
  publisher={Verein zur F{\"o}rderung des Open Access Publizierens in den Quantenwissenschaften}
}

@article{cade2020fermihubbard,
  title={Strategies for solving the {Fermi-Hubbard} model on near-term quantum computers},
  author={Cade, Chris and Mineh, Lana and Montanaro, Ashley and Stanisic, Stasja},
  journal={Physical Review B},
  volume={102},
  number={23},
  pages={235122},
  year={2020},
  publisher={APS}
}

@article{chen2024error,
  title={Error-correcting codes for fermionic quantum simulation},
  author={Chen, Yu-An and Gorshkov, Alexey V and Xu, Yijia},
  journal={SciPost Physics},
  volume={16},
  number={1},
  pages={033},
  year={2024}
}

@article{landahl2021logical,
  title={Logical fermions for fault-tolerant quantum simulation},
  author={Landahl, Andrew J and Morrison, Benjamin CA},
  journal={arXiv preprint arXiv:2110.10280},
  year={2021}
}

@article{jiang2019majorana,
  title={Majorana loop stabilizer codes for error mitigation in fermionic quantum simulations},
  author={Jiang, Zhang and McClean, Jarrod and Babbush, Ryan and Neven, Hartmut},
  journal={Physical Review Applied},
  volume={12},
  number={6},
  pages={064041},
  year={2019},
  publisher={APS}
}

@article{chien2023simulating,
  title={Simulating quantum error mitigation in fermionic encodings},
  author={Chien, Riley W and Setia, Kanav and Bonet-Monroig, Xavier and Steudtner, Mark and Whitfield, James D},
  journal={arXiv preprint arXiv:2303.02270},
  year={2023}
}

@article{clinton2024towards,
  title={Towards near-term quantum simulation of materials},
  author={Clinton, Laura and Cubitt, Toby and Flynn, Brian and Gambetta, Filippo Maria and Klassen, Joel and Montanaro, Ashley and Piddock, Stephen and Santos, Raul A and Sheridan, Evan},
  journal={Nature Communications},
  volume={15},
  number={1},
  pages={211},
  year={2024},
  publisher={Nature Publishing Group UK London}
}

@article{haah2021quantum,
  title={Quantum algorithm for simulating real time evolution of lattice {H}amiltonians},
  author={Haah, Jeongwan and Hastings, Matthew B and Kothari, Robin and Low, Guang Hao},
  journal={SIAM Journal on Computing},
  volume={52},
  number={6},
  pages={FOCS18--250},
  year={2021},
  publisher={SIAM}
}

@Article{AspuruGuzik2005,
	author =	{A. {Aspuru-Guzik} and A. D. Dutoi and P. J. Love and M. {Head-Gordon }},
	title =		{{Simulated Quantum Computation of Molecular Energies}},
	journal =	{Science},
	pages = 	1704,
	year = 		2005,
	doi = 		{10.1126/science.1113479}}

@Article{Kandala2017,
	author =	{A. Kandala and A. Mezzacapo and K. Temme and M. Takita and M. Brink and J. M. Chow and J. M. Gambetta},
	title = 	{Hardware-efficient variational quantum eigensolver for small molecules and quantum magnets},
	journal = 	{Nature},
	volume = 	549,
	pages = 	242,
	year = 		2017,
	doi =		{10.1038/nature23879}}

@Article{Google2020,
	author = 	{{Google AI Quantum and Collaborators} and Frank Arute and Kunal Arya and Ryan Babbush and Dave Bacon and Joseph C. Bardin and Rami Barends and Sergio Boixo and Michael Broughton and Bob B. Buckley and David A. Buell and Brian Burkett and Nicholas Bushnell and Yu Chen and Zijun Chen and Benjamin Chiaro and Roberto Collins and William Courtney and Sean Demura and Andrew Dunsworth and Edward Farhi and Austin Fowler and Brooks Foxen and Craig Gidney and Marissa Giustina and Rob Graff and Steve Habegger and Matthew P. Harrigan and Alan Ho and Sabrina Hong and Trent Huang and William J. Huggins and Lev Ioffe and Sergei V. Isakov and Evan Jeffrey and Zhang Jiang and Cody Jones and Dvir Kafri and Kostyantyn Kechedzhi and Julian Kelly and Seon Kim and Paul V. Klimov and Alexander Korotkov and Fedor Kostritsa and David Landhuis and Pavel Laptev and Mike Lindmark and Erik Lucero and Orion Martin and John M. Martinis and Jarrod R. McClean and Matt McEwen and Anthony Megrant and Xiao Mi and Masoud Mohseni and Wojciech Mruczkiewicz and Josh Mutus and Ofer Naaman and Matthew Neeley and Charles Neill and Hartmut Neven and Murphy Yuezhen Niu and Thomas E. O’Brien and Eric Ostby and Andre Petukhov and Harald Putterman and Chris Quintana and Pedram Roushan and Nicholas C. Rubin and Daniel Sank and Kevin J. Satzinger and Vadim Smelyanskiy and Doug Strain and Kevin J. Sung and Marco Szalay and Tyler Y. Takeshita and Amit Vainsencher and Theodore White and Nathan Wiebe and Z. Jamie Yao and Ping Yeh and Adam Zalcman},
	title = 	"{Hartree-Fock on a superconducting qubit quantum computer}",
	journal = 	{Science},
	volume = 	369,
	pages = 	1084,
	year = 		2020,
	doi = 		{10.1126/science.abb9811}}

@Article{OMalley2016,
	author = 	{O'Malley, P. J. J. and Babbush, R. and Kivlichan, I. D. and Romero, J. and McClean, J. R. and Barends, R. and Kelly, J. and Roushan, P. and Tranter, A. and Ding, N. and Campbell, B. and Chen, Y. and Chen, Z. and Chiaro, B. and Dunsworth, A. and Fowler, A. G. and Jeffrey, E. and Lucero, E. and Megrant, A. and Mutus, J. Y. and Neeley, M. and Neill, C. and Quintana, C. and Sank, D. and Vainsencher, A. and Wenner, J. and White, T. C. and Coveney, P. V. and Love, P. J. and Neven, H. and Aspuru-Guzik, A. and Martinis, J. M.},
	title =		{{Scalable Quantum Simulation of Molecular Energies}},
	journal =	{Phys. Rev. X},
	volume =	6,
	pages =		031007,
	year =		2016,
	doi = 		{10.1103/PhysRevX.6.031007}}

@Article{Hempel2018,
	author =	{Hempel, Cornelius and Maier, Christine and Romero, Jonathan and McClean, Jarrod and Monz, Thomas and Shen, Heng and Jurcevic, Petar and Lanyon, Ben P. and Love, Peter and Babbush, Ryan and Aspuru-Guzik, Al\'an and Blatt, Rainer and Roos, Christian F.},
	title =		{{Quantum Chemistry Calculations on a Trapped-Ion Quantum Simulator}},
	journal =	{Phys. Rev. X},
	volume =	8,
	pages =		031022,
	year = 		2018,
	doi =		{10.1103/PhysRevX.8.031022}}

@article{Colless2018,
	author =	{Colless, J. I. and Ramasesh, V. V. and Dahlen, D. and Blok, M. S. and Kimchi-Schwartz, M. E. and McClean, J. R. and Carter, J. and de Jong, W. A. and Siddiqi, I.},
	title =		{{Computation of Molecular Spectra on a Quantum Processor with an Error-Resilient Algorithm}},
	journal =	{Phys. Rev. X},
	volume = 	8,
	pages =		011021,
	year =		2018,
	doi =		{10.1103/PhysRevX.8.011021}}

@Article{navickas2024experimentalquantumsimulationchemical,
      title={{Experimental Quantum Simulation of Chemical Dynamics}}, 
      author={T. Navickas and R. J. MacDonell and C. H. Valahu and V. C. Olaya-Agudelo and F. Scuccimarra and M. J. Millican and V. G. Matsos and H. L. Nourse and A. D. Rao and M. J. Biercuk and C. Hempel and I. Kassal and T. R. Tan},
      year={2024},
    journal={arXiv preprint arXiv:2409.04044}}

@Article{Miessen2023,
author={A. Miessen and P. J. Ollitrault and F. Tacchino and I. Tavernelli},
title={Quantum algorithms for quantum dynamics},
journal={Nature Comp. Sci.},
year={2023},
volume={3},
pages={25-37},
doi={10.1038/s43588-022-00374-2}}

@article{Ollitrault2020,
  title = {{Nonadiabatic Molecular Quantum Dynamics with Quantum Computers}},
  author = {P. J. Ollitrault and G. Mazzola and I. Tavernelli},
  journal = {Phys. Rev. Lett.},
  volume = {125},
  pages = {260511},
  year = {2020},
  doi = {10.1103/PhysRevLett.125.260511}}

@article{Sawaya2019,
author = {N. P. D. Sawaya and J. Huh},
title = {{Quantum Algorithm for Calculating Molecular Vibronic Spectra}},
journal = {J. Phys. Chem. Lett.},
volume = {10},
pages = {3586-3591},
year = {2019},
doi = {10.1021/acs.jpclett.9b01117}}

@article{Kassal2008,
author = {I. Kassal  and S. P. Jordan  and P. J. Love  and M. Mohseni and A. {Aspuru-Guzik}},
title = {Polynomial-time quantum algorithm for the simulation of chemical dynamics},
journal = {Proc. Nat. Acad. Sci.},
volume = {105},
pages = {18681-18686},
year = {2008},
doi = {10.1073/pnas.0808245105}}

@article{babbush_mean-field_2023,
	title = {Quantum simulation of exact electron dynamics can be more efficient than classical mean-field methods},
	volume = {14},
	issn = {2041-1723},
	url = {https://doi.org/10.1038/s41467-023-39024-0},
	doi = {10.1038/s41467-023-39024-0},
	number = {1},
	journal = {Nature Communications},
	author = {Babbush, Ryan and Huggins, William J. and Berry, Dominic W. and Ung, Shu Fay and Zhao, Andrew and Reichman, David R. and Neven, Hartmut and Baczewski, Andrew D. and Lee, Joonho},
	month = jul,
	year = {2023},
	pages = {4058},
}

@article{babbushQuantumSimulationChemistry2019,
  title = {{Quantum Simulation of Chemistry with Sublinear Scaling in Basis Size}},
  author = {Babbush, Ryan and Berry, Dominic W. and McClean, Jarrod R. and Neven, Hartmut},
  year = {2019},
  month = nov,
  journal = {npj Quantum Information},
  volume = {5},
  number = {1},
  pages = {92},
  issn = {2056-6387},
  doi = {10.1038/s41534-019-0199-y}
}

@article{Shee_qubit_efficient,
  title = {Qubit-efficient encoding scheme for quantum simulations of electronic structure},
  author = {Shee, Yu and Tsai, Pei-Kai and Hong, Cheng-Lin and Cheng, Hao-Chung and Goan, Hsi-Sheng},
  journal = {Phys. Rev. Res.},
  volume = {4},
  issue = {2},
  pages = {023154},
  numpages = {14},
  year = {2022},
  month = {May},
  publisher = {American Physical Society},
  doi = {10.1103/PhysRevResearch.4.023154},
  url = {https://link.aps.org/doi/10.1103/PhysRevResearch.4.023154}
}

@article{su_fault-tolerant_2021,
	title = {Fault-{Tolerant} {Quantum} {Simulations} of {Chemistry} in {First} {Quantization}},
	volume = {2},
	issn = {2691-3399},
	url = {https://link.aps.org/doi/10.1103/PRXQuantum.2.040332},
	doi = {10.1103/PRXQuantum.2.040332},
	language = {en},
	number = {4},
	urldate = {2024-07-19},
	journal = {PRX Quantum},
	author = {Su, Yuan and Berry, Dominic W. and Wiebe, Nathan and Rubin, Nicholas and Babbush, Ryan},
	month = nov,
	year = {2021},
	pages = {040332},
}

@article{babbush_exponentially_2018,
	title = {Exponentially more precise quantum simulation of fermions in the configuration interaction representation},
	volume = {3},
	issn = {2058-9565},
	url = {https://iopscience.iop.org/article/10.1088/2058-9565/aa9463},
	doi = {10.1088/2058-9565/aa9463},
	number = {1},
	urldate = {2024-07-19},
	journal = {Quantum Science and Technology},
	author = {Babbush, Ryan and Berry, Dominic W and Sanders, Yuval R and Kivlichan, Ian D and Scherer, Artur and Wei, Annie Y and Love, Peter J and Aspuru-Guzik, Alán},
	month = jan,
	year = {2018},
	pages = {015006},
}

@article{kirby_secondquant,
  title = {{Second-Quantized Fermionic Operators with Polylogarithmic Qubit and Gate Complexity}},
  author = {Kirby, William and Fuller, Bryce and Hadfield, Charles and Mezzacapo, Antonio},
  journal = {PRX Quantum},
  volume = {3},
  issue = {2},
  pages = {020351},
  numpages = {19},
  year = {2022},
  month = {Jun},
  publisher = {American Physical Society},
  doi = {10.1103/PRXQuantum.3.020351},
  url = {https://link.aps.org/doi/10.1103/PRXQuantum.3.020351}
}

@article{abrams_simulation_1997,
	title = {Simulation of {Many}-{Body} {Fermi} {Systems} on a {Universal} {Quantum} {Computer}},
	volume = {79},
	copyright = {http://link.aps.org/licenses/aps-default-license},
	issn = {0031-9007, 1079-7114},
	url = {https://link.aps.org/doi/10.1103/PhysRevLett.79.2586},
	doi = {10.1103/PhysRevLett.79.2586},
	language = {en},
	number = {13},
	urldate = {2024-07-19},
	journal = {Physical Review Letters},
	author = {Abrams, Daniel S. and Lloyd, Seth},
	month = sep,
	year = {1997},
	pages = {2586--2589},
}

@article{haah2016algebraic,
  title={Algebraic methods for quantum codes on lattices},
  author={Haah, Jeongwan},
  journal={arXiv preprint arXiv:1607.01387},
  year={2016}
}

@article{simkovic2024low,
  title={{Low-Weight High-Distance Error Correcting Fermionic Encodings}},
  author={Simkovic IV, Fedor and Leib, Martin and Pereira, Francisco Revson F},
  journal={arXiv preprint arXiv:2402.15386},
  year={2024}
}

@article{jordan1928about,
	title = {About the {Pauli} exclusion principle},
	volume = {47},
	issn = {1434-6001, 1434-601X},
	url = {http://link.springer.com/10.1007/BF01331938},
	doi = {10.1007/BF01331938},
	language = {de},
	number = {9-10},
	urldate = {2022-07-01},
	journal = {Zeitschrift f{\"u}r Physik},
	author = {Jordan, P. and Wigner, E.},
	month = sep,
	year = {1928},
	pages = {631--651}
}

@article{bringewatt2023parallelization,
  title={Parallelization techniques for quantum simulation of fermionic systems},
  author={Bringewatt, Jacob and Davoudi, Zohreh},
  journal={Quantum},
  volume={7},
  pages={975},
  year={2023},
  publisher={Verein zur F{\"o}rderung des Open Access Publizierens in den Quantenwissenschaften}
}

@ARTICLE{Seeley_2012,title={The {B}ravyi-{K}itaev transformation for quantum computation of electronic structure},year={2012},author={Jacob T. Seeley and Martin J. Richard and Peter J. Love and Peter J. Love},doi={10.1063/1.4768229},pmid={23248989},pmcid={null},mag_id={2037118799},journal={Journal of Chemical Physics}}

@article{miller2024treespilation,
	title={{Treespilation: Architecture- and State-Optimised Fermion-to-Qubit Mappings}}, 
	author={Aaron Miller and Adam Glos and Zoltán Zimborás},
	year={2024},
    journal={arXiv preprint arXiv:2403.03992}
}

@article{miller2023bonsai,
	title = {{Bonsai Algorithm: Grow Your Own Fermion-to-Qubit Mappings}},
	author = {Miller, Aaron and Zimbor\'as, Zolt\'an and Knecht, Stefan and Maniscalco, Sabrina and Garc\'{\i}a-P\'erez, Guillermo},
	journal = {PRX Quantum},
	volume = {4},
	issue = {3},
	pages = {030314},
	numpages = {19},
	year = {2023},
	month = {Aug},
	publisher = {American Physical Society},
	doi = {10.1103/PRXQuantum.4.030314},
	url = {https://link.aps.org/doi/10.1103/PRXQuantum.4.030314}
}

@INPROCEEDINGS{wang2023evermore,
author={Wang, Qingfeng and Cian, Ze-Pei and Li, Ming and Markov, Igor L. and Nam, Yunseong},
booktitle={2023 60th ACM/IEEE Design Automation Conference (DAC)}, 
title={Ever more optimized simulations of fermionic systems on a quantum computer}, 
year={2023},
volume={},
number={},
pages={1-6},
doi={10.1109/DAC56929.2023.10247693}}

@article{steudtner2019quantum,
  title = {Quantum codes for quantum simulation of fermions on a square lattice of qubits},
  author = {Steudtner, Mark and Wehner, Stephanie},
  journal = {Phys. Rev. A},
  volume = {99},
  issue = {2},
  pages = {022308},
  numpages = {32},
  year = {2019},
  month = {Feb},
  publisher = {American Physical Society},
  doi = {10.1103/PhysRevA.99.022308},
  url = {https://link.aps.org/doi/10.1103/PhysRevA.99.022308}
}

@article{berry_improved_2018,
	title = {Improved techniques for preparing eigenstates of fermionic {Hamiltonians}},
	volume = {4},
	copyright = {2018 The Author(s)},
	issn = {2056-6387},
	url = {https://www.nature.com/articles/s41534-018-0071-5},
	doi = {10.1038/s41534-018-0071-5},
	language = {en},
	number = {1},
	urldate = {2024-09-04},
	journal = {npj Quantum Information},
	author = {Berry, Dominic W. and Kieferová, Mária and Scherer, Artur and Sanders, Yuval R. and Low, Guang Hao and Wiebe, Nathan and Gidney, Craig and Babbush, Ryan},
	month = may,
	year = {2018},
	note = {Publisher: Nature Publishing Group},
	pages = {1--7},
}

@article{mcardle2020quantum,
  title={Quantum computational chemistry},
  author={McArdle, Sam and Endo, Suguru and Aspuru-Guzik, Al{\'a}n and Benjamin, Simon C and Yuan, Xiao},
  journal={Reviews of Modern Physics},
  volume={92},
  number={1},
  pages={015003},
  year={2020},
  publisher={APS}
}

@article{cao2019quantum,
  title={Quantum chemistry in the age of quantum computing},
  author={Cao, Yudong and Romero, Jonathan and Olson, Jonathan P and Degroote, Matthias and Johnson, Peter D and Kieferov{\'a}, M{\'a}ria and Kivlichan, Ian D and Menke, Tim and Peropadre, Borja and Sawaya, Nicolas PD and others},
  journal={Chemical reviews},
  volume={119},
  number={19},
  pages={10856--10915},
  year={2019},
  publisher={ACS Publications}
}

@article{motta2022emerging,
  title={Emerging quantum computing algorithms for quantum chemistry},
  author={Motta, Mario and Rice, Julia E},
  journal={Wiley Interdisciplinary Reviews: Computational Molecular Science},
  volume={12},
  number={3},
  pages={e1580},
  year={2022},
  publisher={Wiley Online Library}
}

@article{bauer2020quantum,
  title={Quantum algorithms for quantum chemistry and quantum materials science},
  author={Bauer, Bela and Bravyi, Sergey and Motta, Mario and Chan, Garnet Kin-Lic},
  journal={Chemical Reviews},
  volume={120},
  number={22},
  pages={12685--12717},
  year={2020},
  publisher={ACS Publications}
}

@article{jordan2012quantum,
  title={Quantum algorithms for quantum field theories},
  author={Jordan, Stephen P and Lee, Keith SM and Preskill, John},
  journal={Science},
  volume={336},
  number={6085},
  pages={1130--1133},
  year={2012},
  publisher={American Association for the Advancement of Science}
}

@article{jordan2011quantum,
  title={Quantum computation of scattering in scalar quantum field theories},
  author={Jordan, Stephen P and Lee, Keith SM and Preskill, John},
  journal={arXiv preprint arXiv:1112.4833},
  year={2011}
}

@article{jordan2014quantum,
  title={Quantum algorithms for fermionic quantum field theories},
  author={Jordan, Stephen P and Lee, Keith SM and Preskill, John},
  journal={arXiv preprint arXiv:1404.7115},
  year={2014}
}

@article{jordan2018bqp,
  title={{BQP}-completeness of scattering in scalar quantum field theory},
  author={Jordan, Stephen P and Krovi, Hari and Lee, Keith SM and Preskill, John},
  journal={Quantum},
  volume={2},
  pages={44},
  year={2018},
  publisher={Verein zur F{\"o}rderung des Open Access Publizierens in den Quantenwissenschaften}
}

@article{bauer2023quantum,
  title={Quantum simulation for high-energy physics},
  author={Bauer, Christian W and Davoudi, Zohreh and Balantekin, A Baha and Bhattacharya, Tanmoy and Carena, Marcela and De Jong, Wibe A and Draper, Patrick and El-Khadra, Aida and Gemelke, Nate and Hanada, Masanori and others},
  journal={PRX quantum},
  volume={4},
  number={2},
  pages={027001},
  year={2023},
  publisher={APS}
}

@article{bauer2023quantum_natrevphys,
  title={Quantum simulation of fundamental particles and forces},
  author={Bauer, Christian W and Davoudi, Zohreh and Klco, Natalie and Savage, Martin J},
  journal={Nature Reviews Physics},
  volume={5},
  number={7},
  pages={420--432},
  year={2023},
  publisher={Nature Publishing Group UK London}
}

@article{wan2021exponentially,
  title={Exponentially faster implementations of {S}elect ({H}) for fermionic {H}amiltonians},
  author={Wan, Kianna},
  journal={Quantum},
  volume={5},
  pages={380},
  year={2021},
  publisher={Verein zur F{\"o}rderung des Open Access Publizierens in den Quantenwissenschaften}
}

@article{babbush2018encoding,
  title={Encoding electronic spectra in quantum circuits with linear {T} complexity},
  author={Babbush, Ryan and Gidney, Craig and Berry, Dominic W and Wiebe, Nathan and McClean, Jarrod and Paler, Alexandru and Fowler, Austin and Neven, Hartmut},
  journal={Physical Review X},
  volume={8},
  number={4},
  pages={041015},
  year={2018},
  publisher={APS}
}

@book{gottesman1997stabilizer,
  title={Stabilizer codes and quantum error correction},
  author={Gottesman, Daniel},
  year={1997},
  publisher={California Institute of Technology}
}

@article{zhao2021fermionic,
  title={Fermionic partial tomography via classical shadows},
  author={Zhao, Andrew and Rubin, Nicholas C and Miyake, Akimasa},
  journal={Physical Review Letters},
  volume={127},
  number={11},
  pages={110504},
  year={2021},
  publisher={APS}
}

@article{wan2023matchgate,
  title={Matchgate shadows for fermionic quantum simulation},
  author={Wan, Kianna and Huggins, William J and Lee, Joonho and Babbush, Ryan},
  journal={Communications in Mathematical Physics},
  volume={404},
  number={2},
  pages={629--700},
  year={2023},
  publisher={Springer}
}

@article{low2022classical,
  title={Classical shadows of fermions with particle number symmetry},
  author={Low, Guang Hao},
  journal={arXiv preprint arXiv:2208.08964},
  year={2022}
}

@Inbook{Wick1997,
author="Wick, G. C.
and Wightman, A. S.
and Wigner, E. P.",
title="The Intrinsic Parity of Elementary Particles",
bookTitle="Part I: Particles and Fields. Part II: Foundations of Quantum Mechanics",
year="1997",
publisher="Springer Berlin Heidelberg",
address="Berlin, Heidelberg",
pages="102--106",
isbn="978-3-662-09203-3",
doi="10.1007/978-3-662-09203-3_9",
url="https://doi.org/10.1007/978-3-662-09203-3_9"
}

@article{nigmatullin2024experimental,
  title={{Experimental Demonstration of Break-Even for the Compact Fermionic Encoding}},
  author={Nigmatullin, Ramil and Hemery, Kevin and Ghanem, Khaldoon and Moses, Steven and Gresh, Dan and Siegfried, Peter and Mills, Michael and Gatterman, Thomas and Hewitt, Nathan and Granet, Etienne and others},
  journal={arXiv preprint arXiv:2409.06789},
  year={2024}
}

@article{king2024triply,
  title = {{Triply Efficient Shadow Tomography}},
  author = {King, Robbie and Gosset, David and Kothari, Robin and Babbush, Ryan},
  journal = {PRX Quantum},
  volume = {6},
  issue = {1},
  pages = {010336},
  numpages = {25},
  year = {2025},
  month = {Feb},
  publisher = {American Physical Society},
  doi = {10.1103/PRXQuantum.6.010336},
  url = {https://link.aps.org/doi/10.1103/PRXQuantum.6.010336}
}

@article{babbush2018low,
  title={Low-depth quantum simulation of materials},
  author={Babbush, Ryan and Wiebe, Nathan and McClean, Jarrod and McClain, James and Neven, Hartmut and Chan, Garnet Kin-Lic},
  journal={Physical Review X},
  volume={8},
  number={1},
  pages={011044},
  year={2018},
  publisher={APS}
}

@inproceedings{tan2022analog,
  title={Analog quantum simulation of chemical dynamics with a trapped-ion system},
  author={Tan, Ting Rei and MacDonell, Ryan and Navickas, Tomas and W{\"o}hlers-Reichel, Tim and Rao, Arjun and Biercuk, Michael and Hempel, Cornelius and Kassal, Ivan},
  booktitle={APS Division of Atomic, Molecular and Optical Physics Meeting Abstracts},
  volume={2022},
  pages={C08--005},
  year={2022}
}

@article{liu2024toward,
  title={{Toward Mixed Analog-Digital Quantum Signal Processing: Quantum AD/DA Conversion and the Fourier Transform}},
  author={Liu, Yuan and Martyn, John M and Sinanan-Singh, Jasmine and Smith, Kevin C and Girvin, Steven M and Chuang, Isaac L},
  journal={arXiv preprint arXiv:2408.14729},
  year={2024}
}

@article{crane2024hybrid,
  title={{Hybrid Oscillator-Qubit Quantum Processors: Simulating Fermions, Bosons, and Gauge Fields}},
  author={Crane, Eleanor and Smith, Kevin C and Tomesh, Teague and Eickbusch, Alec and Martyn, John M and K{\"u}hn, Stefan and Funcke, Lena and DeMarco, Michael Austin and Chuang, Isaac L and Wiebe, Nathan and others},
  journal={arXiv preprint arXiv:2409.03747},
  year={2024}
}

@article{harrison2023reducingqubitrequirementjordanwigner,
      title={Reducing the qubit requirement of {Jordan-Wigner} encodings of $N$-mode, $K$-fermion systems from $N$ to $\lceil \log_2 \binom{N}{K} \rceil$}, 
      author={Brent Harrison and Dylan Nelson and Daniel Adamiak and James Whitfield},
      year={2023},
    journal={arXiv preprint arXiv:2211.04501},
}

@article{carolan2024succinctfermiondatastructures,
      title={{Succinct Fermion Data Structures}}, 
      author={Joseph Carolan and Luke Schaeffer},
      year={2024},
        journal={arXiv preprint arXiv:2410.04015}, 
}

@article{Vlasov_2022,
   title={{Clifford Algebras, Spin Groups and Qubit Trees}},
   volume={11},
   ISSN={1314-7374},
   url={http://dx.doi.org/10.12743/quanta.v11i1.199},
   DOI={10.12743/quanta.v11i1.199},
   number={1},
   journal={Quanta},
   publisher={Quanta},
   author={Vlasov, Alexander Yurievich},
   year={2022},
   month=dec, pages={97–114} }

@article{harrison2024sierpinskitrianglefermiontoqubittransform,
      title={{A Sierpinski Triangle Fermion-to-Qubit Transform}}, 
      author={Brent Harrison and Mitchell Chiew and Jason Necaise and Andrew Projansky and Sergii Strelchuk and James D. Whitfield},
      year={2024},
    journal={arXiv preprint arXiv:2409.04348}
}

@article{farhi2000quantum,
  title={Quantum computation by adiabatic evolution},
  author={Farhi, Edward and Goldstone, Jeffrey and Gutmann, Sam and Sipser, Michael},
  journal={arXiv preprint quant-ph/0001106},
  year={2000}
}

@article{albash2018adiabatic,
  title={Adiabatic quantum computation},
  author={Albash, Tameem and Lidar, Daniel A},
  journal={Reviews of Modern Physics},
  volume={90},
  number={1},
  pages={015002},
  year={2018},
  publisher={APS}
}

@article{kato1950adiabatic,
  title={On the adiabatic theorem of quantum mechanics},
  author={Kato, Tosio},
  journal={Journal of the Physical Society of Japan},
  volume={5},
  number={6},
  pages={435--439},
  year={1950},
  publisher={The Physical Society of Japan}
}

@article{childs2021theory,
  title={Theory of trotter error with commutator scaling},
  author={Childs, Andrew M and Su, Yuan and Tran, Minh C and Wiebe, Nathan and Zhu, Shuchen},
  journal={Physical Review X},
  volume={11},
  number={1},
  pages={011020},
  year={2021},
  publisher={APS}
}

@article{childs2019nearly,
  title={Nearly optimal lattice simulation by product formulas},
  author={Childs, Andrew M and Su, Yuan},
  journal={Physical review letters},
  volume={123},
  number={5},
  pages={050503},
  year={2019},
  publisher={APS}
}

@article{kitaev1995quantum,
  title={Quantum measurements and the {A}belian stabilizer problem},
  author={Kitaev, A Yu},
  journal={arXiv preprint quant-ph/9511026},
  year={1995}
}

@article{temme2011quantum,
  title={Quantum metropolis sampling},
  author={Temme, Kristan and Osborne, Tobias J and Vollbrecht, Karl G and Poulin, David and Verstraete, Frank},
  journal={Nature},
  volume={471},
  number={7336},
  pages={87--90},
  year={2011},
  publisher={Nature Publishing Group UK London}
}

@article{chen2023quantum,
  title={Quantum thermal state preparation},
  author={Chen, Chi-Fang and Kastoryano, Michael J and Brand{\~a}o, Fernando GSL and Gily{\'e}n, Andr{\'a}s},
  journal={arXiv preprint arXiv:2303.18224},
  year={2023}
}

@article{chen2023efficient,
  title={An efficient and exact noncommutative quantum {G}ibbs sampler},
  author={Chen, Chi-Fang and Kastoryano, Michael J and Gily{\'e}n, Andr{\'a}s},
  journal={arXiv preprint arXiv:2311.09207},
  year={2023}
}

@article{low2017optimal,
  title={Optimal {H}amiltonian simulation by quantum signal processing},
  author={Low, Guang Hao and Chuang, Isaac L},
  journal={Physical review letters},
  volume={118},
  number={1},
  pages={010501},
  year={2017},
  publisher={APS}
}

@article{low2019hamiltonian,
  title={Hamiltonian simulation by qubitization},
  author={Low, Guang Hao and Chuang, Isaac L},
  journal={Quantum},
  volume={3},
  pages={163},
  year={2019},
  publisher={Verein zur F{\"o}rderung des Open Access Publizierens in den Quantenwissenschaften}
}

@article{berry2009black,
  title={Black-box {H}amiltonian simulation and unitary implementation},
  author={Berry, Dominic W and Childs, Andrew M},
  journal={arXiv preprint arXiv:0910.4157},
  year={2009}
}

@article{childs2012hamiltonian,
  title={Hamiltonian simulation using linear combinations of unitary operations},
  author={Childs, Andrew M and Wiebe, Nathan},
  journal={arXiv preprint arXiv:1202.5822},
  year={2012}
}

@article{huang2020predicting,
  title={Predicting many properties of a quantum system from very few measurements},
  author={Huang, Hsin-Yuan and Kueng, Richard and Preskill, John},
  journal={Nature Physics},
  volume={16},
  number={10},
  pages={1050--1057},
  year={2020},
  publisher={Nature Publishing Group UK London}
}

@article{elben2023randomized,
  title={The randomized measurement toolbox},
  author={Elben, Andreas and Flammia, Steven T and Huang, Hsin-Yuan and Kueng, Richard and Preskill, John and Vermersch, Beno{\^\i}t and Zoller, Peter},
  journal={Nature Reviews Physics},
  volume={5},
  number={1},
  pages={9--24},
  year={2023},
  publisher={Nature Publishing Group UK London}
}

@inproceedings{gilyen2019quantum,
  title={Quantum singular value transformation and beyond: exponential improvements for quantum matrix arithmetics},
  author={Gily{\'e}n, Andr{\'a}s and Su, Yuan and Low, Guang Hao and Wiebe, Nathan},
  booktitle={Proceedings of the 51st Annual ACM SIGACT Symposium on Theory of Computing},
  pages={193--204},
  year={2019}
}

@article{lieb1972finite,
  title={The finite group velocity of quantum spin systems},
  author={Lieb, Elliott H and Robinson, Derek W},
  journal={Communications in mathematical physics},
  volume={28},
  number={3},
  pages={251--257},
  year={1972},
  publisher={Springer}
}

@article{tran2019locality,
  title={Locality and digital quantum simulation of power-law interactions},
  author={Tran, Minh C and Guo, Andrew Y and Su, Yuan and Garrison, James R and Eldredge, Zachary and Foss-Feig, Michael and Childs, Andrew M and Gorshkov, Alexey V},
  journal={Physical Review X},
  volume={9},
  number={3},
  pages={031006},
  year={2019},
  publisher={APS}
}

@article{zhang2023dissipative,
  title={Dissipative quantum {G}ibbs sampling},
  author={Zhang, Daniel and Bosse, Jan Lukas and Cubitt, Toby},
  journal={arXiv preprint arXiv:2304.04526},
  year={2023}
}

@article{cubitt2023dissipative,
  title={Dissipative ground state preparation and the dissipative quantum eigensolver},
  author={Cubitt, Toby S},
  journal={arXiv preprint arXiv:2303.11962},
  year={2023}
}

@incollection{feynman2018simulating,
  title={Simulating physics with computers},
  author={Feynman, Richard P},
  booktitle={Feynman and computation},
  pages={133--153},
  year={2018},
  publisher={cRc Press}
}

@article{manin1980computable,
  title={Computable and uncomputable},
  author={Manin, Yuri},
  journal={Sovetskoye Radio, Moscow},
  volume={128},
  pages={28},
  year={1980}
}

@article{chiew2024ternarytreetransformationsequivalent,
      title={Ternary tree transformations are equivalent to linear encodings of the {F}ock basis}, 
      author={Mitchell Chiew and Brent Harrison and Sergii Strelchuk},
      year={2024},
    journal={arXiv preprint arXiv:2412.07578},
}

@article{o2024ultrafast,
  title={Ultrafast hybrid fermion-to-qubit mapping},
  author={O'Brien, Oliver and Strelchuk, Sergii},
  journal={Physical Review B},
  volume={109},
  number={11},
  pages={115149},
  year={2024},
  publisher={APS}
}

@article{bravyi2006lieb,
  title={Lieb-{R}obinson bounds and the generation of correlations and topological quantum order},
  author={Bravyi, Sergey and Hastings, Matthew B and Verstraete, Frank},
  journal={Physical review letters},
  volume={97},
  number={5},
  pages={050401},
  year={2006},
  publisher={APS}
}

@article{rhodes2024exponential,
    title = {{Exponential Improvements in the Simulation of Lattice Gauge Theories Using Near-Optimal Techniques}},
  author = {Rhodes, Mason L. and Kreshchuk, Michael and Pathak, Shivesh},
  journal = {PRX Quantum},
  volume = {5},
  issue = {4},
  pages = {040347},
  numpages = {32},
  year = {2024},
  month = {Dec},
  publisher = {American Physical Society},
  doi = {10.1103/PRXQuantum.5.040347},
  url = {https://link.aps.org/doi/10.1103/PRXQuantum.5.040347}
}

@article{parelladilme2024reducing,
	title = {{Reducing Entanglement with Physically Inspired Fermion-To-Qubit Mappings}},
	author = {Parella-Dilm\'e, Teodor and Kottmann, Korbinian and Zambrano, Leonardo and Mortimer, Luke and Kottmann, Jakob S. and Ac\'{\i}n, Antonio},
	journal = {PRX Quantum},
	volume = {5},
	issue = {3},
	pages = {030333},
	numpages = {18},
	year = {2024},
	month = {Aug},
	publisher = {American Physical Society},
	doi = {10.1103/PRXQuantum.5.030333},
	url = {https://link.aps.org/doi/10.1103/PRXQuantum.5.030333}
}

@Article{Hubbard1963,
  author =	{J. Hubbard},
  title =		{Electron correlations in narrow energy bands},
  journal =	{Proc. Roy. Soc. London},
  year =		1963,
  volume =	276,
  pages =		238}

@article{reiher2017elucidating,
  title={Elucidating reaction mechanisms on quantum computers},
  author={Reiher, Markus and Wiebe, Nathan and Svore, Krysta M and Wecker, Dave and Troyer, Matthias},
  journal={Proceedings of the national academy of sciences},
  volume={114},
  number={29},
  pages={7555--7560},
  year={2017},
  publisher={National Acad Sciences}
}

@article{rubin2023fault,
  title={Fault-tolerant quantum simulation of materials using {B}loch orbitals},
  author={Rubin, Nicholas C and Berry, Dominic W and Malone, Fionn D and White, Alec F and Khattar, Tanuj and DePrince III, A Eugene and Sicolo, Sabrina and K{\"u}ehn, Michael and Kaicher, Michael and Lee, Joonho and others},
  journal={PRX Quantum},
  volume={4},
  number={4},
  pages={040303},
  year={2023},
  publisher={APS}
}

@article{berry2024quantum,
  title={Quantum simulation of realistic materials in first quantization using non-local pseudopotentials},
  author={Berry, Dominic W and Rubin, Nicholas C and Elnabawy, Ahmed O and Ahlers, Gabriele and DePrince III, A Eugene and Lee, Joonho and Gogolin, Christian and Babbush, Ryan},
  journal={npj Quantum Information},
  volume={10},
  number={1},
  pages={130},
  year={2024},
  publisher={Nature Publishing Group UK London}
}

@article{cobanera2011bond,
  title={The bond-algebraic approach to dualities},
  author={Cobanera, Emilio and Ortiz, Gerardo and Nussinov, Zohar},
  journal={Advances in physics},
  volume={60},
  number={5},
  pages={679--798},
  year={2011},
  publisher={Taylor \& Francis}
}

@article{lee2023evaluating,
  title={Evaluating the evidence for exponential quantum advantage in ground-state quantum chemistry},
  author={Lee, Seunghoon and Lee, Joonho and Zhai, Huanchen and Tong, Yu and Dalzell, Alexander M and Kumar, Ashutosh and Helms, Phillip and Gray, Johnnie and Cui, Zhi-Hao and Liu, Wenyuan and others},
  journal={Nature communications},
  volume={14},
  number={1},
  pages={1952},
  year={2023},
  publisher={Nature Publishing Group UK London}
}

@article{bravyi2017tapering,
  title={Tapering off qubits to simulate fermionic {H}amiltonians},
  author={Bravyi, Sergey and Gambetta, Jay M and Mezzacapo, Antonio and Temme, Kristan},
  journal={arXiv preprint arXiv:1701.08213},
  year={2017}
}

@article{jiang2020optimal,
  title={Optimal fermion-to-qubit mapping via ternary trees with applications to reduced quantum states learning},
  author={Jiang, Zhang and Kalev, Amir and Mruczkiewicz, Wojciech and Neven, Hartmut},
  journal={Quantum},
  volume={4},
  pages={276},
  year={2020},
  publisher={Verein zur F{\"o}rderung des Open Access Publizierens in den Quantenwissenschaften}
}

@article{havlivcek2017operator,
  title={Operator locality in the quantum simulation of fermionic models},
  author={Havl{\'\i}{\v{c}}ek, Vojt{\v{e}}ch and Troyer, Matthias and Whitfield, James D},
  journal={Physical Review A},
  volume={95},
  number={3},
  pages={032332},
  year={2017},
  publisher={APS}
}

@article{li2022higher,
  title={Higher-dimensional {J}ordan-{W}igner transformation and auxiliary {M}ajorana fermions},
  author={Li, Kangle and Po, Hoi Chun},
  journal={Physical Review B},
  volume={106},
  number={11},
  pages={115109},
  year={2022},
  publisher={APS}
}

@article{ortiz2001quantum,
  title={Quantum algorithms for fermionic simulations},
  author={Ortiz, G and Gubernatis, JE and Knill, E and Laflamme, R},
  journal={Physical Review A},
  volume={64},
  number={2},
  pages={022319},
  year={2001},
  publisher={APS}
}

@article{evered2025probing,
  title={Probing topological matter and fermion dynamics on a neutral-atom quantum computer},
  author={Evered, Simon J and Kalinowski, Marcin and Geim, Alexandra A and Manovitz, Tom and Bluvstein, Dolev and Li, Sophie H and Maskara, Nishad and Zhou, Hengyun and Ebadi, Sepehr and Xu, Muqing and others},
  journal={arXiv preprint arXiv:2501.18554},
  year={2025}
}

@article{yu2025cliffordcircuitbasedheuristic,
      title={Clifford circuit based heuristic optimization of fermion-to-qubit mappings}, 
      author={Jeffery Yu and Yuan Liu and Sho Sugiura and Troy Van Voorhis and Sina Zeytinoğlu},
      year={2025},
    journal={arXiv preprint arXiv:2502.11933}, 
}

@article{setia2020reducing,
  title={Reducing qubit requirements for quantum simulations using molecular point group symmetries},
  author={Setia, Kanav and Chen, Richard and Rice, Julia E and Mezzacapo, Antonio and Pistoia, Marco and Whitfield, James D},
  journal={Journal of Chemical Theory and Computation},
  volume={16},
  number={10},
  pages={6091--6097},
  year={2020},
  publisher={ACS Publications}
}

@article{lanyon2010towards,
  title={Towards quantum chemistry on a quantum computer},
  author={Lanyon, Benjamin P and Whitfield, James D and Gillett, Geoff G and Goggin, Michael E and Almeida, Marcelo P and Kassal, Ivan and Biamonte, Jacob D and Mohseni, Masoud and Powell, Ben J and Barbieri, Marco and others},
  journal={Nature chemistry},
  volume={2},
  number={2},
  pages={106--111},
  year={2010},
  publisher={Nature Publishing Group}
}

@article{wecker2014gate,
  title={Gate-count estimates for performing quantum chemistry on small quantum computers},
  author={Wecker, Dave and Bauer, Bela and Clark, Bryan K and Hastings, Matthew B and Troyer, Matthias},
  journal={Physical Review A},
  volume={90},
  number={2},
  pages={022305},
  year={2014},
  publisher={APS}
}

@article{lloyd1996universal,
  title={Universal quantum simulators},
  author={Lloyd, Seth},
  journal={Science},
  volume={273},
  number={5278},
  pages={1073--1078},
  year={1996},
  publisher={American Association for the Advancement of Science}
}

@article{suzuki1991general,
  title={General theory of fractal path integrals with applications to many-body theories and statistical physics},
  author={Suzuki, Masuo},
  journal={Journal of mathematical physics},
  volume={32},
  number={2},
  pages={400--407},
  year={1991},
  publisher={American Institute of Physics}
}

@article{kogut1975hamiltonian,
  title={Hamiltonian formulation of {W}ilson's lattice gauge theories},
  author={Kogut, John and Susskind, Leonard},
  journal={Physical Review D},
  volume={11},
  number={2},
  pages={395},
  year={1975},
  publisher={APS}
}

@article{tong2022provably,
  title={Provably accurate simulation of gauge theories and bosonic systems},
  author={Tong, Yu and Albert, Victor V and McClean, Jarrod R and Preskill, John and Su, Yuan},
  journal={Quantum},
  volume={6},
  pages={816},
  year={2022},
  publisher={Verein zur F{\"o}rderung des Open Access Publizierens in den Quantenwissenschaften}
}

@article{watson2023quantum,
  title={Quantum algorithms for simulating nuclear effective field theories},
  author={Watson, James D and Bringewatt, Jacob and Shaw, Alexander F and Childs, Andrew M and Gorshkov, Alexey V and Davoudi, Zohreh},
  journal={arXiv preprint arXiv:2312.05344},
  year={2023}
}

@article{whitfield2011simulation,
  title={Simulation of electronic structure {H}amiltonians using quantum computers},
  author={Whitfield, James D and Biamonte, Jacob and Aspuru-Guzik, Al{\'a}n},
  journal={Molecular Physics},
  volume={109},
  number={5},
  pages={735--750},
  year={2011},
  publisher={Taylor \& Francis}
}

@article{kuhn2014quantum,
  title={Quantum simulation of the {S}chwinger model: {A} study of feasibility},
  author={K{\"u}hn, Stefan and Cirac, J Ignacio and Ba{\~n}uls, Mari-Carmen},
  journal={Physical Review A},
  volume={90},
  number={4},
  pages={042305},
  year={2014},
  publisher={APS}
}

@article{farrell2024scalable,
  title={Scalable circuits for preparing ground states on digital quantum computers: The {S}chwinger model vacuum on 100 qubits},
  author={Farrell, Roland C and Illa, Marc and Ciavarella, Anthony N and Savage, Martin J},
  journal={PRX Quantum},
  volume={5},
  number={2},
  pages={020315},
  year={2024},
  publisher={APS}
}

@article{schwinger1962gauge,
  title={Gauge invariance and mass. {II}},
  author={Schwinger, Julian},
  journal={Physical Review},
  volume={128},
  number={5},
  pages={2425},
  year={1962},
  publisher={APS}
}

@article{lamm2019general,
  title={General methods for digital quantum simulation of gauge theories},
  author={Lamm, Henry and Lawrence, Scott and Yamauchi, Yukari and (NuQS Collaboration)},
  journal={Physical Review D},
  volume={100},
  number={3},
  pages={034518},
  year={2019},
  publisher={APS}
}

@article{alexandru2019gluon,
  title={Gluon field digitization for quantum computers},
  author={Alexandru, Andrei and Bedaque, Paulo F and Harmalkar, Siddhartha and Lamm, Henry and Lawrence, Scott and Warrington, Neill C and (NuQS Collaboration)},
  journal={Physical Review D},
  volume={100},
  number={11},
  pages={114501},
  year={2019},
  publisher={APS}
}

@article{davoudi2021search,
  title={Search for efficient formulations for {H}amiltonian simulation of non-{A}belian lattice gauge theories},
  author={Davoudi, Zohreh and Raychowdhury, Indrakshi and Shaw, Andrew},
  journal={Physical Review D},
  volume={104},
  number={7},
  pages={074505},
  year={2021},
  publisher={APS}
}

@article{lamm2024block,
  title={Block encodings of discrete subgroups on a quantum computer},
  author={Lamm, Henry and Li, Ying-Ying and Shu, Jing and Wang, Yi-Lin and Xu, Bin},
  journal={Physical Review D},
  volume={110},
  number={5},
  pages={054505},
  year={2024},
  publisher={APS}
}

@article{kan2021lattice,
  title={Lattice quantum chromodynamics and electrodynamics on a universal quantum computer},
  author={Kan, Angus and Nam, Yunseong},
  journal={arXiv preprint arXiv:2107.12769},
  year={2021}
}

@article{nielsen1981absence,
  title={Absence of neutrinos on a lattice: ({I}). {P}roof by homotopy theory},
  author={Nielsen, Holger Bech and Ninomiya, Masao},
  journal={Nuclear Physics B},
  volume={185},
  number={1},
  pages={20--40},
  year={1981},
  publisher={Elsevier}
}

@article{wecker2015solving,
  title={Solving strongly correlated electron models on a quantum computer},
  author={Wecker, Dave and Hastings, Matthew B and Wiebe, Nathan and Clark, Bryan K and Nayak, Chetan and Troyer, Matthias},
  journal={Physical Review A},
  volume={92},
  number={6},
  pages={062318},
  year={2015},
  publisher={APS}
}

@article{bauer2016hybrid,
  title={Hybrid quantum-classical approach to correlated materials},
  author={Bauer, Bela and Wecker, Dave and Millis, Andrew J and Hastings, Matthew B and Troyer, Matthias},
  journal={Physical Review X},
  volume={6},
  number={3},
  pages={031045},
  year={2016},
  publisher={APS}
}

@article{bluvstein2024logical,
  title={Logical quantum processor based on reconfigurable atom arrays},
  author={Bluvstein, Dolev and Evered, Simon J and Geim, Alexandra A and Li, Sophie H and Zhou, Hengyun and Manovitz, Tom and Ebadi, Sepehr and Cain, Madelyn and Kalinowski, Marcin and Hangleiter, Dominik and others},
  journal={Nature},
  volume={626},
  number={7997},
  pages={58--65},
  year={2024},
  publisher={Nature Publishing Group UK London}
}

@article{google2023suppressing,
  title={Suppressing quantum errors by scaling a surface code logical qubit},
  author={Acharya, Rajeev and Aleiner, Igor and Allen, Richard and Andersen, Trond I and Ansmann, Markus and Arute, Frank and Arya, Kunal and Asfaw, Abraham and Atalaya, Juan and Babbush, Ryan and others},
  journal={Nature},
  volume={614},
  number={7949},
  pages={676--681},
  year={2023},
  publisher={Nature Publishing Group UK London}
}

@article{paetznick2024demonstration,
  title={Demonstration of logical qubits and repeated error correction with better-than-physical error rates},
  author={Paetznick, A and da Silva, MP and Ryan-Anderson, C and Bello-Rivas, JM and Campora III, JP and Chernoguzov, A and Dreiling, JM and Foltz, C and Frachon, F and Gaebler, JP and others},
  journal={arXiv preprint arXiv:2404.02280},
  year={2024}
}

@article{peruzzo2014variational,
  title={A variational eigenvalue solver on a photonic quantum processor},
  author={Peruzzo, Alberto and McClean, Jarrod and Shadbolt, Peter and Yung, Man-Hong and Zhou, Xiao-Qi and Love, Peter J and Aspuru-Guzik, Al{\'a}n and O’brien, Jeremy L},
  journal={Nature communications},
  volume={5},
  number={1},
  pages={4213},
  year={2014},
  publisher={Nature Publishing Group UK London}
}

@article{arute2020observation,
  title={Observation of separated dynamics of charge and spin in the {Fermi--Hubbard} model},
  author={Arute, Frank and Arya, Kunal and Babbush, Ryan and Bacon, Dave and Bardin, Joseph C and Barends, Rami and Bengtsson, Andreas and Boixo, Sergio and Broughton, Michael and Buckley, Bob B and others},
  journal={arXiv preprint arXiv:2010.07965},
  year={2020}
}

@article{low2025fast,
  title={Fast quantum simulation of electronic structure by spectrum amplification},
  author={Low, Guang Hao and King, Robbie and Berry, Dominic W and Han, Qiushi and DePrince III, A Eugene and White, Alec and Babbush, Ryan and Somma, Rolando D and Rubin, Nicholas C},
  journal={arXiv preprint arXiv:2502.15882},
  year={2025}
}

@article{motlagh2024generalized,
  title={Generalized quantum signal processing},
  author={Motlagh, Danial and Wiebe, Nathan},
  journal={PRX Quantum},
  volume={5},
  number={2},
  pages={020368},
  year={2024},
  publisher={APS}
}

@article{su2021fault,
  title={Fault-tolerant quantum simulations of chemistry in first quantization},
  author={Su, Yuan and Berry, Dominic W and Wiebe, Nathan and Rubin, Nicholas and Babbush, Ryan},
  journal={PRX Quantum},
  volume={2},
  number={4},
  pages={040332},
  year={2021},
  publisher={APS}
}

@article{Wang2021resourceoptimized,
  doi = {10.22331/q-2021-07-26-509},
  url = {https://doi.org/10.22331/q-2021-07-26-509},
  title = {Resource-{O}ptimized {F}ermionic {L}ocal-{H}amiltonian {S}imulation on a {Q}uantum {C}omputer for {Q}uantum {C}hemistry},
  author = {Wang, Qingfeng and Li, Ming and Monroe, Christopher and Nam, Yunseong},
  journal = {{Quantum}},
  issn = {2521-327X},
  publisher = {{Verein zur F{\"{o}}rderung des Open Access Publizierens in den Quantenwissenschaften}},
  volume = {5},
  pages = {509},
  month = jul,
  year = {2021}
}

\end{document}